\documentclass[print]{aa}  
\usepackage{natbib}
\bibpunct{(}{)}{;}{a}{}{,}
\usepackage{graphicx}
\usepackage{txfonts}
\usepackage{color}

\begin{document} 

%-----------------------------------------------------------------------------------------------------------
% TITLE
%-----------------------------------------------------------------------------------------------------------
	\title{Self-consistent atmosphere modeling with cloud formation\\ for low-mass stars and exoplanets}
	\titlerunning{Atmosphere modeling with cloud formation}
	\author{Diana Juncher\inst{1}\thanks{\email{diana@nbi.dk}}, Uffe G. J{\o}rgensen\inst{1} \and Christiane Helling\inst{2}}
	\institute{
		Niels Bohr Institute \& Centre for Star and Planet Formation, University of Copenhagen, {\O}ster Voldgade 5, 1350 Copenhagen, DK
		\and
            	Centre for Exoplanet Science, SUPA, School of Physics and Astronomy, University of St Andrews, North Haugh, St Andrews, Fife KY16 9SS, UK
	}
	\date{}
	\abstract	
  	% context
	{Low-mass stars and extrasolar planets have ultra-cool atmospheres where a rich chemistry occurs and clouds form. 
	The increasing amount of spectroscopic observations for extrasolar planets requires self-consistent model atmosphere simulations 
	to consistently include the formation processes that determine cloud formation and their feedback onto the atmosphere.}
  	% aims
   	{Complement the {\sc Marcs} model atmosphere suit with simulations applicable to low-mass stars and exoplanets in 
	preparation of E-ELT, JWST, PLATO and other upcoming facilities.}
  	% methods
	{The {\sc Marcs} code calculates stellar atmosphere models, providing self-consistent solutions of the radiative transfer 
	and the atmospheric structure and chemistry. We combine {\sc Marcs} with a kinetic model that describes cloud formation 
	in ultra-cool atmospheres (seed formation, growth/ evaporation, gravitational settling, convective mixing, element depletion).}
  	% results
	{We present a small grid of self-consistently calculated atmosphere models for $T_ \text{eff} = 2000 - 3000$ K with solar 
	initial abundances and $\log(g) = 4.5$. Cloud formation in stellar and sub-stellar atmospheres appears for $T_\text{eff} < 
	2700$ K and has a significant effect on the structure and the spectrum of the atmosphere for $T_\text{eff} < 2400$ K. 
	We have compared the synthetic spectra of our models with observed spectra and found that they fit the spectra of mid 
	to late type M-dwarfs and early type L-dwarfs well. {\color{black}The geometrical extension of the atmospheres (at $\tau=1$) changes 
	with wavelength resulting in a flux variation of $\sim$10\%. This translates into a change in geometrical extension of 
	the atmosphere of about 50 km, which is the quantitative basis for exoplanetary transit spectroscopy}. 
	We also test {\sc Drift-Marcs} for an example exoplanet and demonstrate 
	that our simulations reproduce the Spitzer observations for WASP-19b rather well for $T_{\rm eff}=2600$K, $\log(g)={\color{black}3.2}$ and 
	solar abundances. {\color{black} Our model points at an exoplanet with a deep cloud-free atmosphere with a substantial day-night 
	energy transport and no temperature inversion}.}
	% conclusions
	{}
	\maketitle

%-----------------------------------------------------------------------------------------------------------
% INTRODUCTION
%-----------------------------------------------------------------------------------------------------------
\section{Introduction}\label{s:intro}
The atmospheres of late type M-dwarf stars and brown dwarfs -
collectively referred to as ultra cool dwarfs - and planets have low
enough temperatures for clouds to form. Cloud formation increases the
total atmospheric opacity but also affects the local gas phase by
element depletion. A strong effect on the structure of their
atmospheres results, and hence self-consistent inclusion of cloud 
formation is critical for inferring correctly
the physical structure and chemical composition of these objects from
observed spectra. The same physics and considerations apply to the
atmospheres of the bulk of known exoplanets, and the present paper is
therefore our first paper in a series of planned papers to describe
self-consistent modeling of exoplanetary atmospheres as a tool for
interpreting coming high-quality spectra of exoplanets that will
become available with the next generation instruments during the
coming years.

The presence of cloud formation in ultra cool dwarf atmospheres was
first proposed by \citet{lunine86} from the comparison of
temperature-pressure profiles of brown dwarf atmosphere models with
the condensation curves of refractory materials such as iron,
sodium-aluminum silicates, and magnesium silicates. A decade later,
\citet{tsuji96b} presented the first cloud modeling results for brown
dwarfs, showing how the high opacity of dust particles can produce a
noticeably effect in the observed spectrum. \citet{tsuji96b} also
suggested that cloud formation should be considered for all objects
with $T_\text{eff} < 2800$ K and should therefore also be included in
the models of late type M dwarf atmospheres.

Modeling cloud formation is a complex problem involving 
different coupled processes that depend on a wide range of physical
and chemical parameters. Many of the early models of cloudy
atmospheres (e.g. \citet{rossow78}, \citet{lewis69},
\citet{carlson88}, \citet{lunine86}, \citet{burrows89}, and
\citet{tsuji96b}) were able to reproduce basic features of ultra cool
dwarfs by simply turning on or off the opacity of dust in the
atmosphere at its chemical equilibrium temperature-pressure
location. Over the years the models have grown more detailed and more
realistic, and today several independent groups are working on complex
models that represent clouds in atmosphere models using different
strategies. Some are based on practical considerations
\citep{tsuji01,barman11,burrows06}, while others are inspired by
measurements of the atmospheres of the planets in our Solar system
\citep{allard01,cooper03}, terrestrial cloud formation
\citep{ackerman01}, or kinetic dust-formation modeling in asymptotic
giant branch stars \citep{helling01,woitke03,woitke04}. A detailed
comparison between a selection of these can be found in
\citet{helling08b}.

In this paper we present an extension of the {\sc Marcs} code
\citep{gu1971,jorgensen92,gustafsson08,vaneck16} that has so far been
used extensively for modeling atmospheres of cool stars
\citep{lambert86,plez92,ar1997}, including abundance analysis (e.g.
\cite{black1995,mat2013,nis2014,hill2016,siq2016}), H$_2$O detections
\citep{ry2002,ar2002}, microdiamonds in carbon stars \citep{and1995},
and instrument calibration \citep{dec2003,dec2007}. {\sc Marcs} has also
been used to study cool, helium-rich white dwarfs \citep{jor2000}, R
Coronae Borealis stars \citep{asp2000}, and to determine fundamental
properties of GRB progenitors \citep{gro2013}. While the
radiative-transfer treatment of {\sc Marcs} has inspired time-dependent
carbon-rich models for dust-forming AGB stars \citep{ho1998}, the
lower mass counterpart, i.e. late type M-dwarfs and brown dwarfs with 
clouds, has not been
addressed by the {\sc Marcs} community so far. This paper presents {\sc Marcs}
model atmosphere simulations which include a detailed 
modeling of cloud formation, by self-consistently solving the 
radiative transfer and gas-phase chemistry in the scheme of {\sc marcs}
together with the seed formation, growth/evaporation
of cloud particles, element conservation and gravitational
settling in the scheme of {\sc drift}. 
In this way the radiative and chemical feedback on the atmosphere due to 
cloud formation is fully taken into account.
Section~\ref{s:approach} summarizes our approach, including tables of
input properties.  We present our results for a grid of {\sc Drift-Marcs} model 
atmosphere simulations applicable to
solar-metallicity M-dwarfs and brown dwarfs ($T_{\rm
  eff}=2000-3000$ K, $\log(g)=4.5$; Section~\ref{s:results}). 
{\color{black} These models represent an extension of the {\sc Marcs} code 
with respect to the updated gas-phase opacity data and the modeling 
of cloud formation. They also offer a new alternative to the 
{\sc Drift-Phoenix} models.}
 %These models represent an update of the previous {\sc Drift-Phoenix} models
%with respect to the gas-phase opacity data taken into account. 
We compare 
the synthetic spectra resulting from our
atmosphere simulations with observed spectra of mid- to late-type M-dwarfs,
early to mid-type L-dwarfs, and an example giant gas planet WASP-19b in
Section~\ref{s:synspec}. Section~\ref{s:disc} discusses 
the effect of porosity in cloud
particles. Appendix~\ref{s:detailed_spectrum} provides additional details about
the gas species contributing to the synthetic spectra.

%-----------------------------------------------------------------------------------------------------------
% APPROACH
%-----------------------------------------------------------------------------------------------------------
\section{Approach}\label{s:approach}
Two well-tested codes are combined to enable hands-on atmosphere
simulations for ultra-cool, cloud-forming objects. {\sc Drift}, the
cloud formation module, has been applied to investigate cloud
structures in brown dwarfs and extrasolar planets from first
principles (e.g. \citet{helling08,street15,hell2016}). {\sc Marcs} has
been applied to a large number of atmosphere problems
(Section~\ref{s:intro}). We follow a similar strategy as in
\cite{witte09} in combining the two codes. In the following, we
provide a summery of the two codes, the opacity data used, and the
methodology for running the combined codes.

\subsection{\sc Marcs}
\paragraph{The code:}
The {\sc Marcs} code was introduced in the early 1970s by
\citet{gustafsson75} and has since then been developed in step with
the advancement of computer power and available physical data. The
most recent general grid of {\sc Marcs} models was published by
\citet{gustafsson08} and contains about 50,000 state-of-the-art
stellar atmosphere models extending from late A-type to early M-type
stars - from dwarfs to supergiants - for varying metallicities and
C/O-ratios. This version of {\sc Marcs} is very similar to our
version, and details of the implementation of hydrostatic equilibrium,
radiative transfer, convection and mixing length can be found in
\citet{gustafsson08}. For the equilibrium calculations we use a
version of Tsuji's program \citep{tsuji64} implemented by
\citet{helling96}, and updated further for the present work.

\paragraph{Input data:}
The chemical equilibrium calculations in {\sc Marcs} are based on
38 atoms and 210 molecules (see Appendix \ref{s:atoms_and_molecules}). We have adopted the chemical composition
of the Sun as reported by \citet{grevesse07} for all our models.
For the atoms and ions we use the internal partition function data
from \citet{irwin81}
%and \cite{popovas16}
to calculate the equilibrium constants. For the molecules we use the
Gibbs free energy data from \citet{tsuji73,burrows99,burrows05} to
calculate the equilibrium constants.

\begin{table}
\caption{Sources of data for continuum opacities. "b-f" and "f-f" denote bound-free 
and free-free processes, respectively. CIA stands for collision induced absorption.}             
\label{table:continuum-abs}      
\centering          
\begin{tabular}{l l l}
\hline\hline       
Ion 							& Process		& Reference\\
\hline                    
H$^-$						& b-f			& \citet{doughty66} \\
H$^-$						& f-f			& \citet{doughty66b} \\
H{\scriptsize{I}}					& b-f, f-f		& \citet{karzas61} \\
H{\scriptsize{I}}+H{\scriptsize{I}}	& CIA		& \citet{doyle68} \\
H$_2^-$						& f-f			& \citet{somerville64} \\
H$_2^+$						& f-f			& \citet{mihalas65} \\
He$^-$						& f-f			& \citet{somerville65, john67} \\
He{\scriptsize{I}}				& f-f	& \citet{peach70} \\
C{\scriptsize{I}}, 				& f-f	& \citet{peach70} \\
Mg{\scriptsize{I}}				& f-f	& \citet{peach70} \\
Al{\scriptsize{I}},				& f-f	& \citet{peach70} \\
Si{\scriptsize{I}}					& f-f	& \citet{peach70} \\
e$^-$						& scattering	& \citet{mihalas78} \\
H{\scriptsize{I}}					& scattering	& Dalgarno, quoted by \citet{kurucz70} \\
\hline                  
\end{tabular}
\end{table}

We calculate the continuum absorption for about a dozen ions, 
electron scattering and Rayleigh scattering by H{\scriptsize{I}}
(Table~\ref{table:continuum-abs}).
The line opacities for atoms and ions were updated by
\citet{popovas14} with atomic line data from VALD-3
\citep{kupka11}. The line opacities for molecules were updated to
include the 24 molecules and molecular pairs listed in Table~\ref{table:molecule-abs}.  
We sampled all line opacities using the Opacity Sampling method with a
resolution of $R$ = $\lambda/\Delta \lambda$ = 20,000 in the
wavelength range $0.125 - 25\ \mu$m.

\begin{table}
\small
\caption{Molecular line transitions and their sources.}      
\label{table:molecule-abs}      
\centering          
\begin{tabular}{l l l}
\hline\hline       
Molecule 							& Transitions		& Reference\\
\hline    
{\bf{Hydrides}} & &\\
LiH		& vib-rot				& \citet{coppola11} \\
MgH		& vib-rot				& \citet{yadin12} \\
		& A-X, B'-X			& \citet{gharibnezhad13} \\
SiH		& A-X				& \citet{kurucz11} \\
CaH		& vib-rot				& \citet{yadin12} \\
		& A-X, B-X, C-X, D-X, E-X	& \citet{weck03} \\
TiH		& A-X, B-X			& \citet{burrows05} \\
CrH		& A-X				& \citet{burrows02} \\
FeH		& F-X				& \citet{wende10} \\
CH		& vib-rot, A-X, B-X, C-X	& \citet{masseron14} \\
NH		& vib-rot				& \citet{brooke14b} \\
		& A-X, A-C			& \citet{kurucz11} \\
OH		& vib-rot, A-X			& \citet{kurucz11} \\	
		&					& \\
{\bf{Oxides}} & & \\
SiO		& vib-rot				& \citet{barton13} \\
		& A-X, E-X			& \citet{kurucz11} \\
TiO		& A-X, B-X, C-X, E-X		& \citet{schwenke98} \\
		& c-a, b-a, b-d, f-a		& \\
VO		& A-X, B-X, C-X		& \citet{kurucz11} \\
ZrO		& B-A, B-X, C-X, E-A		& \citet{plez03} \\
		& b-a, d-a, e-a, f-a		& \\
CO		& vib-rot, A-X			& \citet{kurucz11} \\
NO		& vib-rot				& \citet{rothman10} \\
H$_2$O	& vib-rot				& \citet{jorgensen01} \\
		&					& \\
{\bf{Other}} & & \\
H$_2$, HD	& vib-rot, quad, B-X, C-X	& \citet{kurucz11} \\
C$_2$	& A-X, b-a, E-A			& \citet{kurucz11} \\
		& d-a				& \citet{brooke13} \\
CN		& vib-rot, A-X, B-X		& \citet{brooke14} \\
CO$_2$	& vib-rot				& \citet{rothman10} \\
HCN		& vib-rot				& \citet{harris06} \\
		&					& \citet{harris08} \\
H$_2$-H$_2$	& CIA    			& \citet{borysow2001} \\
H$_2$-He   	& CIA    			& \citet{jor2000} \\
\hline                  
\end{tabular}
\end{table}

{\color{black}As described in \citet{gustafsson08}, the convection 
in {\sc Marcs} is handled using the mixing length method, 
where the convective energy flux can be calculated as 
a function of the mixing length $l$.
The value of $l$ is based on empirical calibrations of stellar 
interior models and is thus not theoretically derived. It is often 
expressed as a product of the mixing length parameter $\alpha$ 
and the scale height. For cool stars and brown dwarfs 
$\alpha \approx 2$ \citep{ludwig02} and this is the value we 
adopt for our models.}

\subsection{\sc Drift}
\paragraph{The code:}
The {\sc Drift} code models cloud formation by considering each of
the involved physical and chemical processes in detail. The formation
of seed particles and the subsequent growth or evaporation of dust
grains are describe by modified classical nucleation theory and the
moment method, respectively \citep{gail88,dominik93,lee2015}. The
initial model equations where extended to describe the
growth/evaporation of particles of mixed material composition as
required in particular for oxygen-rich atmospheres
\citep{helling06,helling08}. This is coupled to the effects of
gravitational settling, convective mixing and element depletion via a
system of partial differential equations \citep{woitke03,woitke04,helling06}.

{\color{black}The convection in ultra cool dwarfs allow for the upwards transport and 
subsequent diffusion of the non-depleted gas from the interior of the 
dwarf. This convective mixing can be extended into the upper, 
radiative atmosphere via overshooting, thereby facilitating a 
replenishment of the depleted gas above the cloud base, maintaining
 the dust cycle. The {\sc Drift} code models overshooting 
by assuming an exponential decrease of the mass exchange frequency 
above the radiative zone (Equation 9 in \citet{woitke04}, 
with $\beta$=2.2 and $\tau_\text{mix}^\text{min} = 2/(H_pv_c$)).}

We consider seven growth species (TiO$_2$[s], MgSiO$_4$[s],
SiO$_2$[s], Fe[s], Al$_2$O$_3$[s], MgO[s] and MgSiO$_3$[s]) to make
this initial implementation as simple as possible. We include 32
chemical surface reactions which is a subset of reactions of
\cite{helling08} for the respective materials. For TiO$_2$ we use
the data from \citet{woitke03} to calculate the saturation vapor
pressure at different temperatures. For the remaining condensates we
use the data from \citet{sharp90}.

\paragraph{Dust opacity:}
{\sc Drift} calculates the vertical distribution of the clouds as
well as the size and composition of their cloud particles, but to assess
how the opacity of the clouds affect the structure we also need to
calculate the absorption and scattering of the dust grains.

From the information provided by {\sc Drift} about a specific cloud
particle size and the volume of each of its components, we can use the
Bruggeman Equations \citep{bruggeman35} to calculate its effective
index of refraction, assuming that the dust grain is compact and its
components are randomly mixed. This allow us to treat the dust grain
as a homogeneous particle, the properties of its components combining
to generate effective properties of the whole particle itself.

Because the size of the dust grains are typically of the same order as the wavelength of the starlight, we cannot 
use the Rayleigh or geometrical approximations to describe how they interact with 
the light. Instead we have to use full Mie Theory \citep{mie08,bohren83} for a complete 
description of how electromagnetic plane weaves are absorbed and scattered by 
homogeneous spherical particles. This, of course, also requires the assumption that 
the dust grains are spherical. 

\paragraph{Input Data:}
The sources of the optical constants used to calculate the effective
index of refraction of the mixed dust particles are given in Table
\ref{table:optical-constants}. Most of the data covers the wavelength
range $1.25 - 25\ \mu$m, only the data for Al$_2$O$_3$[s] and
MgSiO$_3$[s] had to be extrapolated down to the lowest considered
wavelength.  We did this by freezing the optical constants from the
first known wavelength points.

\begin{table}
\caption{References for $n$ and $k$ optical constants of the condensates.}      
\label{table:optical-constants}      
\centering          
\begin{tabular}{l l l}
\hline\hline       
Solid species 			& Reference\\
\hline                    
TiO$_2$[s]			& Ribarsky in \citet{palik85} \\
MgSiO$_4$[s]			& \citet{jager03} \\
SiO$_2$[s]			& \citet{posch03} \\
Fe[s]					& \citet{posch03} \\
Al$_2$O$_3$[s]		& \citet{zeidler13} \\
MgO[s]				& Roessler \& Huffman in \citet{palik85} \\
MgSiO$_3$[s]			& \citet{dorschner95} \\
\hline                  
\end{tabular}
\end{table}

\subsection{Merging {\sc Marcs} with Drift}
In order to calculate the details of the cloud layers in an
atmosphere, {\sc Drift} needs information about the ($T_{\rm g}$,
P$_{\rm g}$)-structure, chemical composition and convection of the
atmosphere. Similarly, {\sc Marcs} needs information about the size
and composition of the cloud particles as well as the depletion of
elements to calculate the effects of clouds in the atmosphere. We
manage this data exchange between {\sc Marcs} and {\sc Drift}
through input and output files containing the information listed in
Table \ref{table:interface}.
\begin{table}
\caption{The data exchanged between {\sc Marcs} and {\sc Drift}.}      
\label{table:interface}      
\centering          
\begin{tabular}{r l   |   r l}
\hline\hline       
{\sc Marcs} to {\sc Drift}&				&				& {\sc Drift} to {\sc Marcs} \\
\hline     
layer height				& $z$			& $a(z)$			& average grain size \\
gas temperature			& $T(z)$			& 				& \\
gas pressure				& $P(z)$			& $V_i(z)$			& average grain \\
gas density				& $\rho(z)$		& 				& volume fractions \\
gravitational acceleration		& $g(z)$			&				& \\
convection velocity			& $v_c(z)$		& $\epsilon_i(z)$	& depleted element  \\
mixing length parameter		& $l$				&				& abundances\\
initial element abundances	& $\epsilon_i^0$	&				& \\
\hline                  
\end{tabular}
\end{table}

\paragraph{Changes to the {\sc Marcs} code:}
In previous versions of the {\sc Marcs} code the element abundances 
have been considered constant throughout the atmosphere. Since diffusion 
of atoms is a very slow process that only becomes dominant in stars hotter 
than $T_\text{eff} \approx 11,500$ K \citep{hui-bon-hoa00}, this is usually 
an excellent approximation and especially so for late type stars, where the 
deep convective envelopes will keep the gas well mixed. However, in 
ultra cool dwarfs the dust formation will cause a depletion of elements in 
the top layers where the dust grains form, and a corresponding augmentation 
of elements in the layers where the dust grains evaporates. We have therefore 
expanded the initial one-dimensional array containing element abundances 
with an extra dimension to account for their depth dependence.

{\color{black}The coolest models of \cite{gustafsson08} start at 
a Rosseland optical depth of $\log(\tau) = -5$, ends at 
$\log(\tau) = 2$, and have a resolution of 
$\Delta\log(\tau) = 0.2$. This is appropriate for cloud free 
models, and extending the atmospheres or increasing the resolution 
would have little effect on the computed model. But since cloud formation 
can set in at much lower optical depths than $\log(\tau)=-5$ we found 
it necessary to extend our models out to $\log(\tau)=-10$. 
Furthermore, we also increased the resolution in the upper layers to 
$\Delta\log(\tau) = 0.15$ to accommodate the potentially rapid 
changes over short distances cloud formation can cause.}

\paragraph{Changes to the {\sc Drift} code:}
The changes to the {\sc Drift} code were minimal, consisting only of the 
addition of a new routine to handle the communication with {\sc Marcs}.

\paragraph{Running {\sc Drift}-{\sc Marcs}:}
While {\sc Marcs} needs to know the element depletion and dust opacity 
before it can solve the radiative transfer equation, {\sc Drift} needs to know 
the convection speed which is calculated by {\sc Marcs} as it solves 
the radiative transfer equation. At a first glance it seems like we are in a deadlock, 
but the solution is actually quite simple. If we start with a dust free model of 
$T_\text{eff} \approx 3000$ K and then proceed to gradually lower the effective 
temperature, iterating through {\sc Marcs} and {\sc Drift} for each step, 
the data exchange files will be updated in sync with the increasing dust formation. 
For this to work the change in the effective temperature between each step had to  
be relatively small, about  $T_\text{eff} = 10-50$ K depending on the impact of the 
dust formation.

The dust free version of {\sc Marcs} will keep iterating over a model until the 
temperature corrections in all layers are below a given value, usually  $T \le 2\,$K. However, 
if we allow {\sc Marcs} to fulfill this convergence criterion every time we run 
{\sc Drift}, we can easily end up in a endless loop with no convergence in sight. 
When {\sc Drift} adds a layer of dust to the atmosphere, {\sc Marcs} will heat 
the layers as a reaction to the increased opacity. In response, {\sc Drift} will then 
reduce the amount of dust as the higher temperatures impede the dust formation. 
{\sc Marcs} will of course react to the decreased opacity by cooling the layers again, 
and we are thus back where we started - or even further away! To avoid this, we only let 
{\sc Marcs} iterate once between each call to {\sc Drift}, and we limit the 
temperature correction to half of what the code suggests. This way we stop the overheating 
of the atmosphere and allow the dust formation to react to the temperature change before it 
becomes too large. When the temperature correction is below  $T \le 10$ K we consider the 
cloud layer stable and let {\sc Marcs} converge fully without calling {\sc Drift} again. 
%Include flow chart?

%-----------------------------------------------------------------------------------------------------------
% RESULTS
%-----------------------------------------------------------------------------------------------------------
\section{Results}\label{s:results}
We have created a small grid of models for late type M-dwarfs and
early L-type brown dwarfs with effective temperatures of $T_\text{eff}
= 2000-3000$ K in steps of $T = 100$ K. They all have solar initial
abundances and a surface gravity of $\log(g) = 4.5$. The specifics of
these atmosphere models are discussed in the following.

\subsection{Atmosphere models}\label{s:atm}   
\begin{figure}
\hspace*{-0.5cm}
\includegraphics[width=10cm]{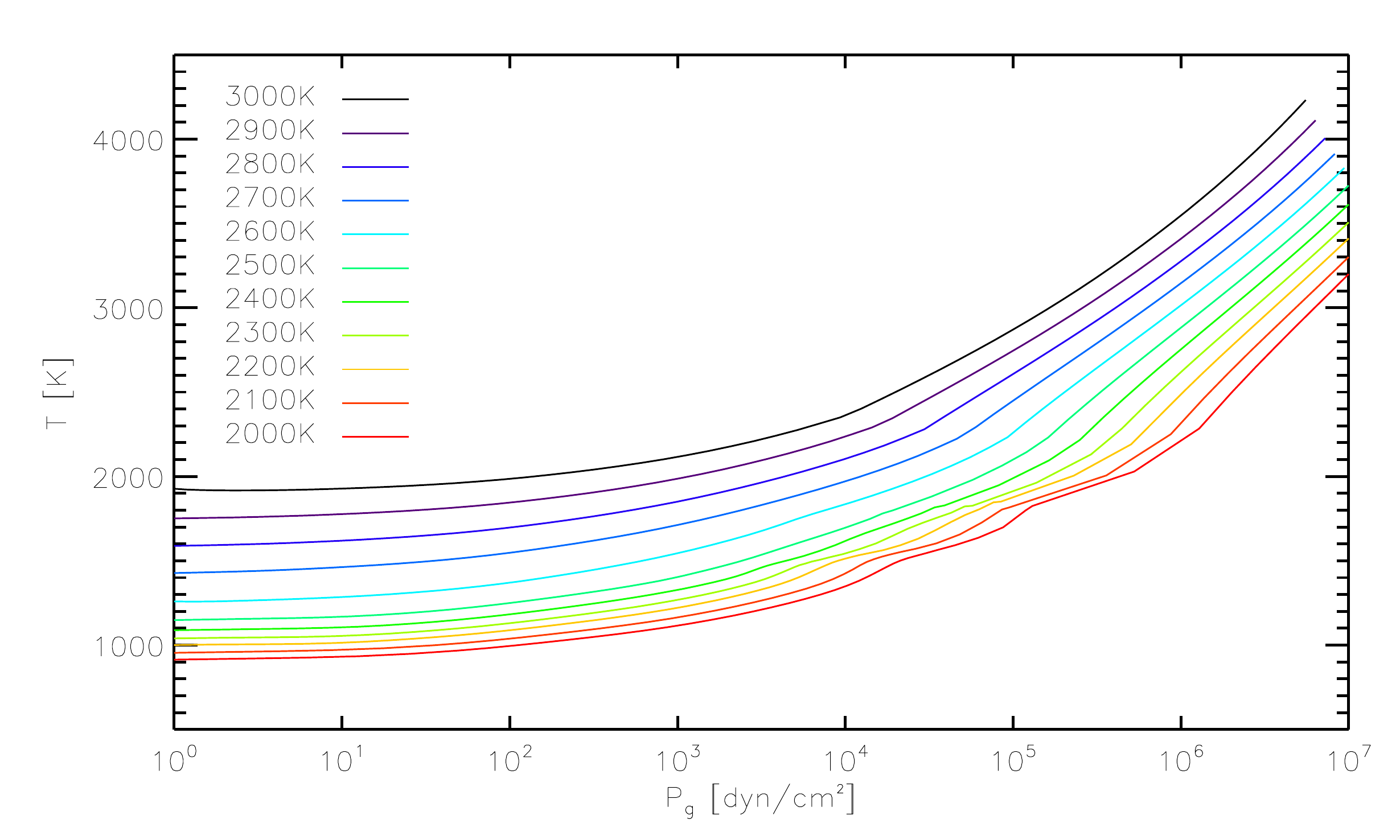}
\caption{The $T$-$P_g$ profiles for our grid of models with varying effective temperatures, $\log(g) = 4.5$ and solar initial abundances.}
\label{fig:temperature-pressure}
\end{figure}

%\subsubsection{Structure}
We present the temperature-pressure profiles of our models in Figure
\ref{fig:temperature-pressure}.  {\color{black}Convection sets in at around $P_g >
10^5 \text{ dyn/cm}^{-1}$ and is the predominant mode of energy
transport in the bottom layers of the atmosphere. In the upper layers 
the temperature gradient is very shallow}.

\begin{figure}
\hspace*{-0.5cm}
\includegraphics[width=10cm]{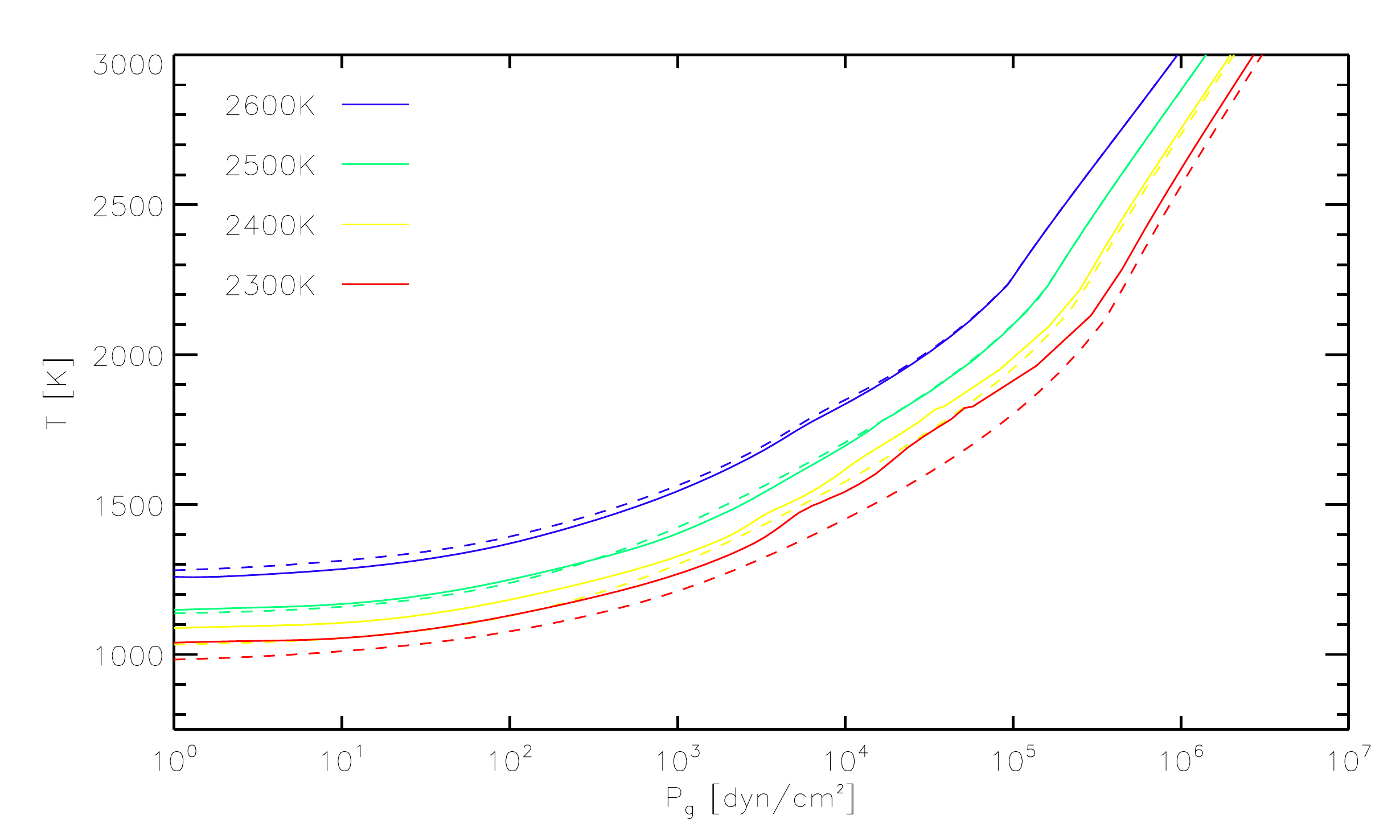}
\caption{The $T$-$P_g$ profiles for our grid of models with clouds (full drawn lines) compared to cloud free models (dashed lines). All models have $\log(g) = 4.5$ and solar initial abundances.}
\label{fig:temperature-pressure2}
\end{figure}

In Figure \ref{fig:temperature-pressure2} we compare the temperature-pressure
profiles of our cloud forming models with models where the cloud forming 
has been switched off. For $T_\text{eff} < 2700$ K the temperature in the upper layers of the
atmosphere is low enough for cloud formation to take place, but the
effect is so small in the beginning, that it barely affects the
structure of the model. At $T_\text{eff} = 2600$ K the amount of dust
formation has increased enough to cause a cooling effect in the outer
layers. This happens because {\color{black}the depletion of the gas phase elements 
that are now bound in dust grains leads to a depletion of the gas phase 
molecules that are made up of those specific elements. Although these 
molecules are a small fraction of the overall number of molecules, 
some of them are important absorbers and their depletion significantly 
reduces the opacity of the atmosphere}. As
long as the clouds are not substantial enough for their own opacity to
compensate for the decreased molecular opacity, 
the affected layers will cool a little. At $T_\text{eff} =
2600$ K the upper layers cool about $10-20$ K, at $T_\text{eff} =
2500$ K the increase in cloud opacity more or less balances the
decrease in molecular opacity, and at $T_\text{eff} = 2400$ K there is
a clear heating of the upper layers caused by cloud formation. For
successively cooler models the amount of heating in the upper layers
increases correspondingly, and in the atmospheres of the coolest
models the back-warming becomes more pronounced. The growing 
irregularities in the temperature at $P_\text{g} = 10^4-10^6$ dyn/cm$^2$ 
coincide with the lower and densest part of the forming clouds (see Figure \ref{fig:dust-info}). 
%Unfortunately, we can 
%only compare with cloud free models down to $T_\text{eff} = 2300$ K. 
%Below this temperature the cloud free models become physically unrealistic 
%and the chemical equations break down without the added opacity of the clouds.

There is a general tendency for the temperature irregularities to shift
downwards for decreasing effective temperatures. This can be explained
by a combination of the thermal stability temperature moving downward,
plus the withdrawal of the convection zone and a lower velocity of the
convective cells, which makes the element replenishment less effective
and causes the clouds to sink down a little into the atmosphere.

\subsubsection{Cloud particle details}
\begin{figure}
\centering
\includegraphics[width=\hsize]{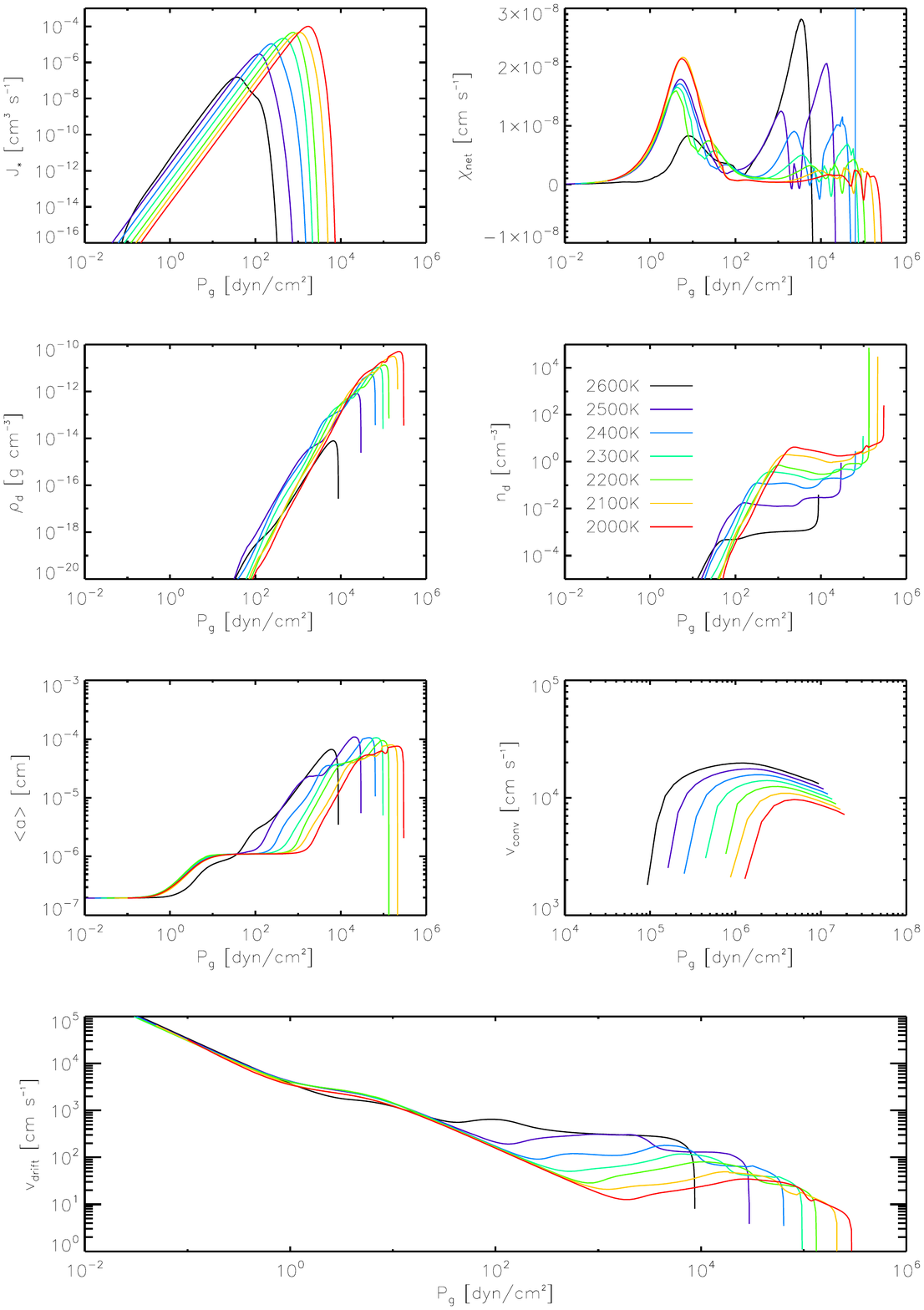}
\caption{The nucleation rate $J_*$, net growth rate $\chi_\text{net}$, mass density 
$\rho_\text{d}$ and number density $n_\text{d}$ of the dust grains, the mean grain 
size <$a$>, the convective velocity $v_\text{conv}$, and the drift velocity $v_\text{d}$ as a function of gas pressure.
Color coding is indicated in panel 4.}
\label{fig:dust-info}
\end{figure}

\begin{figure}
\centering
\includegraphics[width=\hsize]{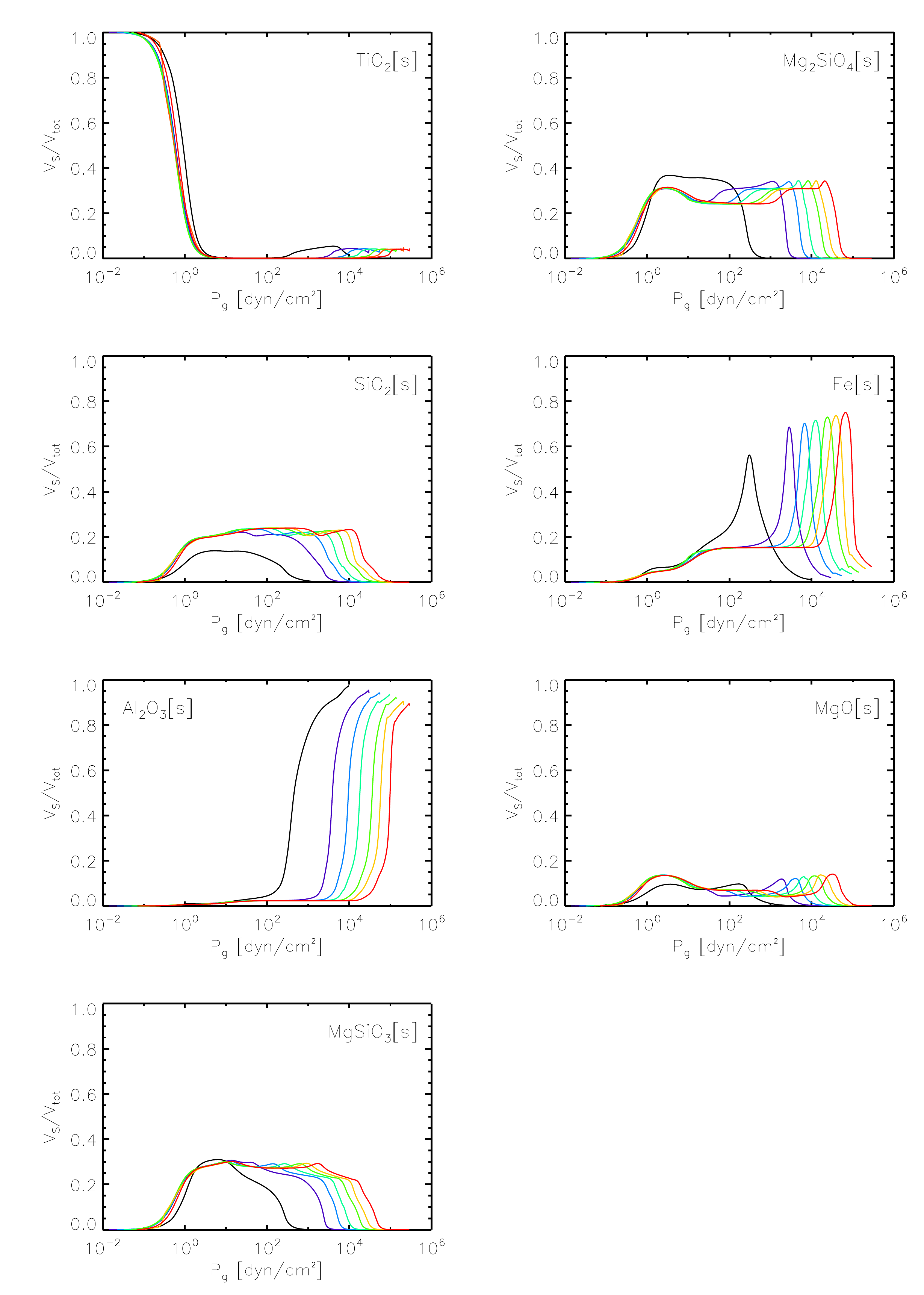}
\caption{The volume fraction of each of the seven solid species in a dust grain as 
a function of gas pressure. All fractions sum up to unity. The color coding is the same as in 
Figure\,\ref{fig:dust-info}.}
\label{fig:dust-volume}
\end{figure}

Figures \ref{fig:dust-info} and \ref{fig:dust-volume} present a more 
detailed view of how the different processes involved in cloud
formation depend on and react to each other as we move down through 
the atmosphere, as well as how the size and composition of the cloud particles
changes in response. 


Starting at the top of the atmosphere and moving down, the nucleation rate $J_*$ 
rises quickly due to increasing collisional rates as the density increases. When a distinct local temperature 
$T_\theta \approx 1300 $ K is exceeded, the nucleation rate drops to zero very fast. 
Consequently, the peak of the nucleation rate reaches deeper into the atmosphere 
the cooler the effective temperature of the model is.
The nucleation rate causes a rise in the number density of dust grains $n_\text{d}$, and 
the peak coincides with the first rapid increase in the number of dust particles. 
% ChH: This is wrong:
%Since  the nucleation rate is much larger than the net growth rate at these heights, the dust 
%grains are almost entirely composed of TiO$_2$[s].

After the nucleation rate peaks the number density flattens out until
it sharply increases again at the bottom of the cloud layers as the
result of gravitational settling and cloud particle accumulation
before complete evaporation. In the middle of the cloud layer, the cloud particle mass density
$\rho_\text{d}$ keeps increasing while the number density does not,
showing that while the nucleation of new cloud particles have
stopped, the already existing ones fall into deeper layers and are still growing larger. This
coincides with the growing volume fraction of cloud particles other than
TiO$_2$[s]. The silicates Mg$_2$SiO$_4$[s], MgSiO$_3$[s] and
SiO$_2$[s] are the first to condense on the seed particles, quickly
followed by MgO[s] and Fe[s] and then finally Al$_2$O$_3$[s].

At the bottom of the cloud layers the cloud particles evaporate at the high temperatures, 
causing a rapid decrease in their mass density and a drop in average particle size.

The net growth rate is $\chi_\text{net} > 0$ when the grains are growing and 
$\chi_\text{net} < 0$ when the grains are evaporating. The first growth period begins 
when we are far enough down in the atmosphere for the solids to effectively condense 
on the nucleation particles. It peaks before the nucleation rate indicating that it 
depends more on the amount of available surface area than on the formation of new 
small particles. The mean grain size <$a$> is determined by the net growth rate, and 
the first and second increase in the mean grain size happens in sync with the first 
and second period of growth. Near the bottom of the cloud layers the net growth rate 
and mean grain size rapidly drops as the cloud particles completely evaporate. 
The fluctuations in the net growth rate is due to the different solid species evaporating 
at different temperatures. 

The drift velocity is initially decreasing as the gas density - and
therefore the friction - increases. The decreasing ends when the
second period of growth sets in, as the larger cloud particles can
more easily overcome the friction with the surrounding gas, because
the downward accelerating force is proportional to grain size cubed
(i.e.\ the grain mass), while the upward restoring force (the
friction) is proportional to grain size squared.

\begin{figure}
\centering
\includegraphics[width=\hsize]{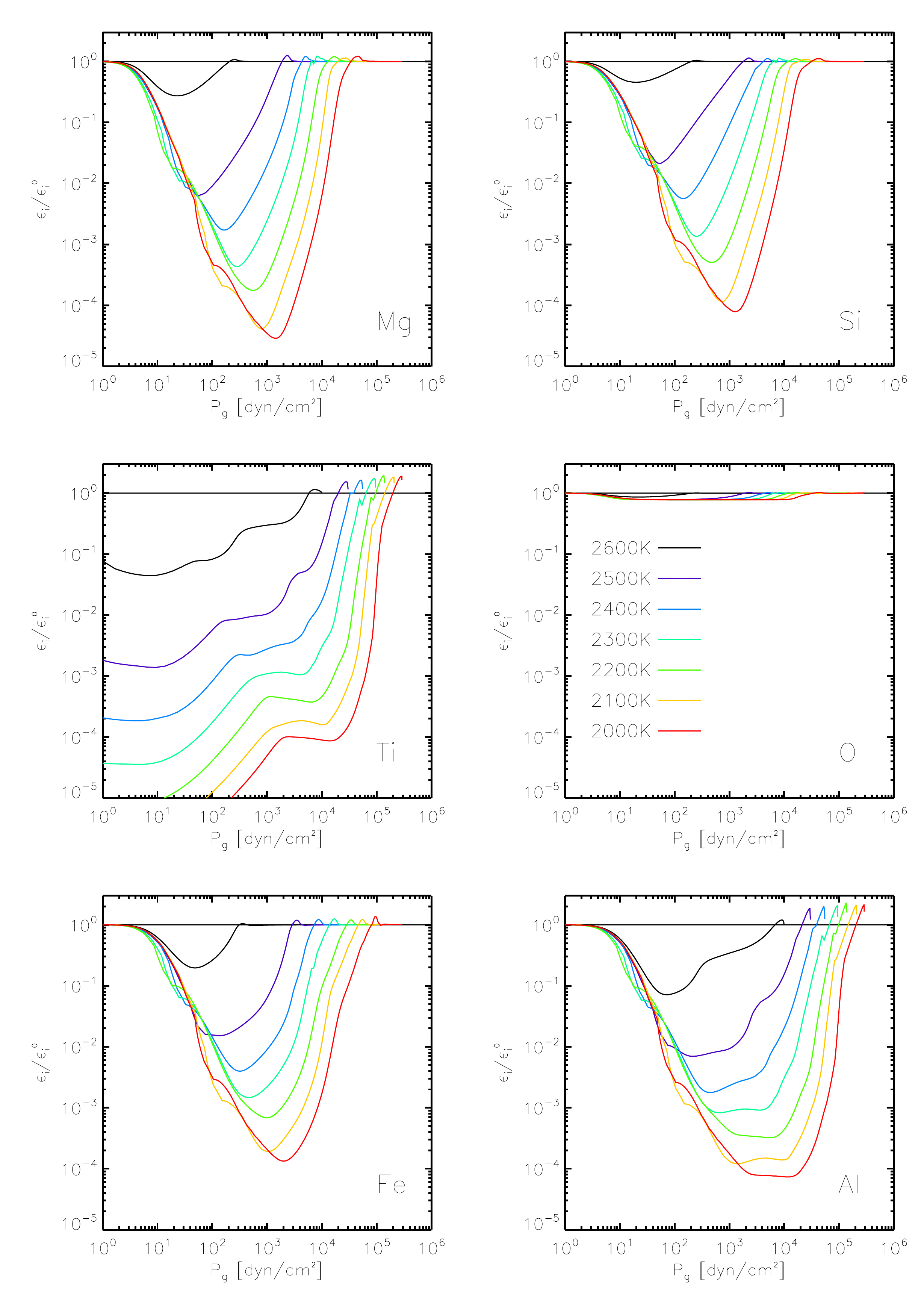}
\caption{Relative element depletion in the gas phase as a function of depth.
Color coding is the same as in Figures \ref{fig:dust-info} and \ref{fig:dust-volume} 
and is also indicated with the legend in panel 4.}
\label{fig:element-depletion}
\end{figure}

\subsubsection{Element depletion}
The cloud particles are formed from the elements Mg, Si, Ti, O, Fe and
Al in the present model. Figure \ref{fig:element-depletion} shows how their abundances in
the gas phase change as a function of atmospheric depth as they are bound in cloud
particles. In general, the more rare elements are stronger
depleted. While the large abundance of O is barely affected by the
cloud formation, the other elements are clearly depleted in the cloud
forming regions. Since we use TiO$_2$[s] as the seed particle, the
relatively small abundance of Ti is strongly depleted in the upper
layers.  The element depletion of the remaining elements sets in a
little later when there is available seed particles for them to
condense on. The depletion is largest where the nucleation peaks
(compare Figures \ref{fig:element-depletion} and \ref{fig:dust-info})
and then decreases as the cloud particles reach the lower warmer layers
and start to evaporate. Because the elements rain out with the cloud
particles, we see an overabundance of condensable elements right below
the cloud base, which is then transported upward with the gas
convection. This will result in an increase of the corresponding gas opacity species.

\subsubsection{Cloud regions}
Based on the considerations in the previous section we can identify different regions 
within the clouds that each have their own characteristics and dominant processes. 
The formation, growth, and evaporation of the cloud particles in a gas of a given
chemical composition is a function of temperature as well as gas pressure, and 
therefore to be understood as ''a race" between the changing values of these two variables. 
Decreasing the temperature will enhance cloud particle formation, as will increasing gas pressure.
The changing conditions for a cloud particles during its movement down through the photosphere,
with its increasing temperature and increasing gas density, is therefore determined
by the ratio of these two variables.

\begin{figure}
\centering
\includegraphics[width=\hsize]{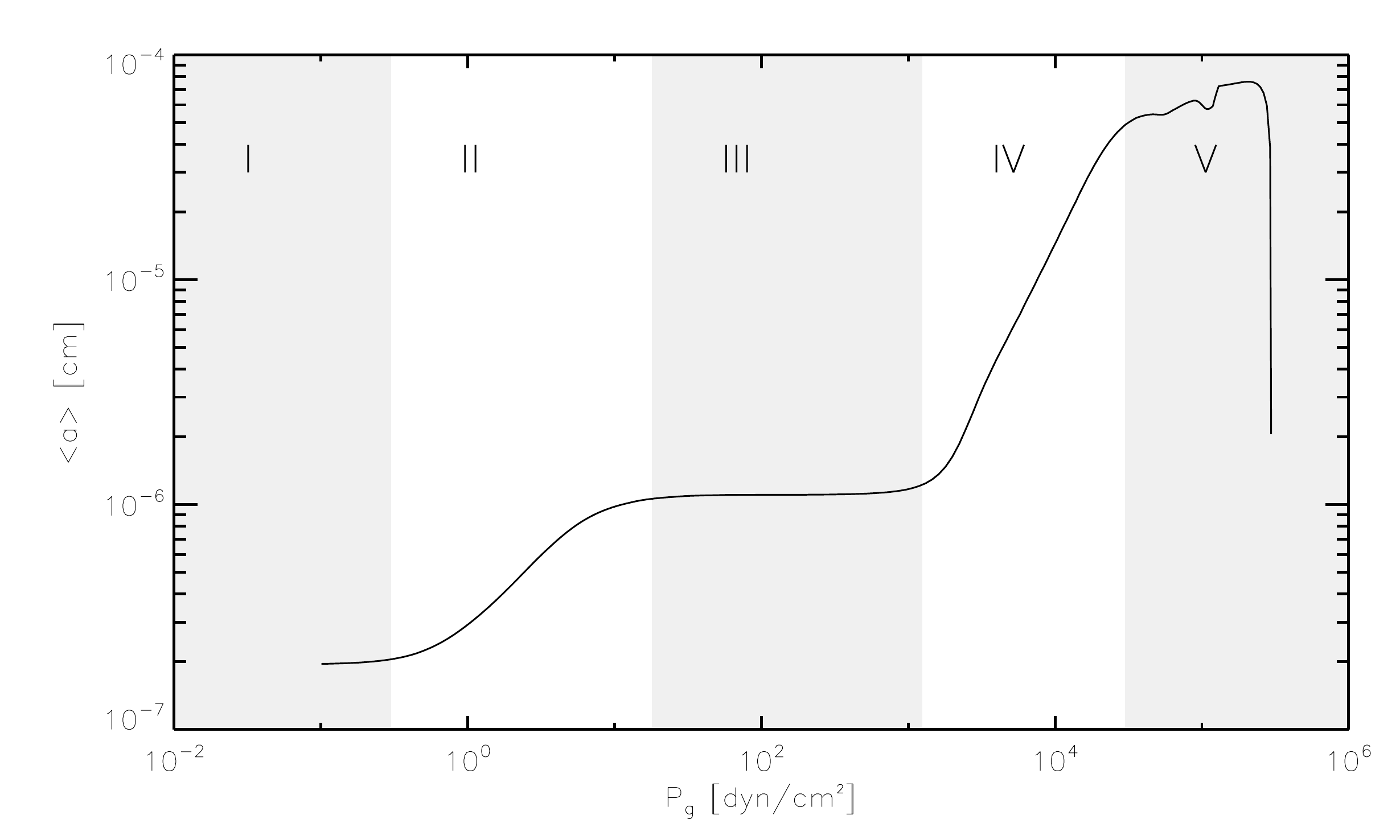}
\caption{The five regions of the clouds: I nucleation, II first growth, III drift, IV second growth, V evaporation.}
\label{fig:regions}
\end{figure}

With respect 
to the grain size distribution, we can divide the clouds into five distinct zones 
based on how the mean size of a cloud particles changes as we move from the top 
to the base of the clouds. The five regions are illustrated for a model with 
$T_\text{eff} = 2000$ K, $\log(g) = 4.5$ and [M/H] = 0.0 in Figure \ref{fig:regions}. 
Similar to \cite{woitke04}, they can be characterized as follows:

\begin{enumerate}
\item {\bf{Nucleation}}\\ At the top of the cloud layers the
  nucleation of gas molecules is the dominant process, and the gas
  phase is therefore highly depleted in Ti relative to all other elements.

\item {\bf{First growth}}\\ As the cloud particles fall down into the
  atmospheric layers, the increasing density and element
  replenishment allow for a growing number of possible surface
  reactions on the small seed particles, and the cloud particles increase
  considerably in size.  As a result, the gas becomes more and more
  depleted in the elements that make up the cloud particles.  The rate of
  newly forming seed particles still increases in this region, but it
  is the rapid growth that has the dominant effect on the average 
  cloud particle size <$a$>.

\item {\bf{Drift}}\\ The increasing density of the gas combined with
  the increasing <$a$> causes the decent of the  cloud particles to slow,
  which reduces the collision rate between the cloud particles and the gas
  molecules. This will decrease the growth rate. In the same region
  the nucleation rate peaks and the average cloud particle size remains
  constant, as the impeded growth of the large cloud particles is
  compensated for by the rapid formation of new small grains.

\item {\bf{Second growth}}\\
When the nucleation rate suddenly drops, the growth in grain size is no longer balanced by the 
formation of small cloud particles, and the mean grain size therefore increases rapidly again. 
This ends the decrease of the drift velocity which remains more or less constant until the 
grains start evaporating. This and the still increasing density allow for an increase in the 
net growth rate.

\item {\bf{Evaporation}} \\
In the lowest layers of the cloud particles  start to evaporate as the temperature reaches 
the monomerization energies of the different solids, and the mean grain size decreases, dropping 
very fast as the last cloud particles evaporate at the cloud base.
\end{enumerate}

We note that this stratification prevails as long as the hydrodynamic
time scales are longer than any of the time scales representing cloud
formation processes.

%-----------------------------------------------------------------------------------------------------------
% SYNTHETIC SPECTRA
%-----------------------------------------------------------------------------------------------------------
\section{Synthetic spectra}\label{s:synspec}
Detailed studies of the many complex physical and chemical processes 
that take place in our model atmospheres are an important part of 
understanding and developing our theories of stars, but at the end of the 
day it is only the light that leaves the atmosphere, the emitted spectrum, 
that provides us with a way to directly compare our stellar models with 
observations of real stars.

\subsection{Gas and cloud opacities}
\begin{figure}
\includegraphics[width=\hsize]{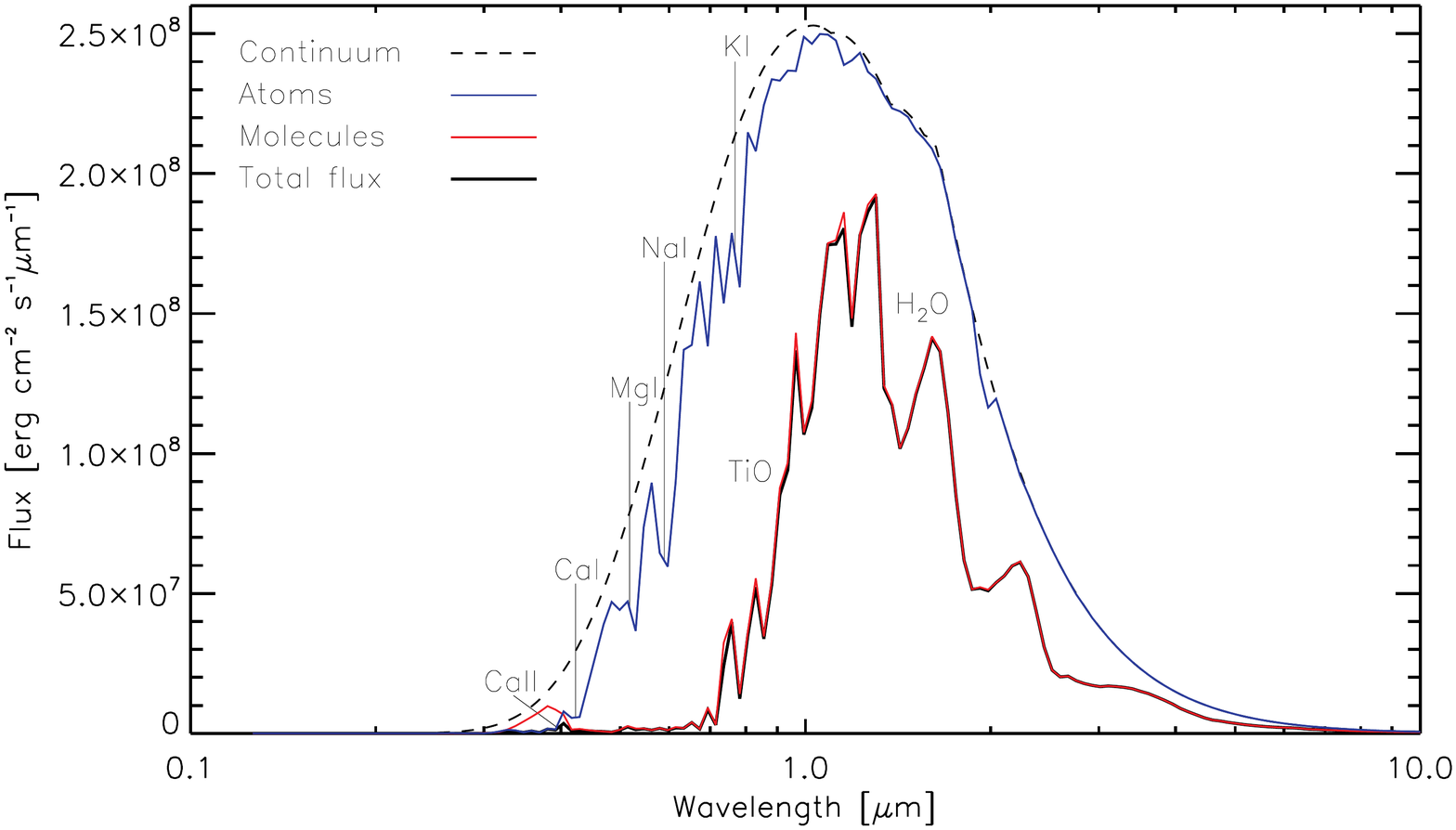}
{\ }\\*[-1.7cm]
\caption{Spectral contributions of gas opacity sources for a cloud-free {\sc Marcs}-model atmosphere for $T_{\rm eff}$=2500 K, $\log(g)=4.5$, [M/H]=0.0.} 
\label{fig:specfree}
\end{figure}

 Figure~\ref{fig:specfree} illustrates the impact of atomic  
 (blue) and molecular (red) line absorption on the spectrum of a dust
 free model with $T_{\rm eff}$ = 2500 K, $\log(g) = 4.5$ and [M/H] =
 0. At such low effective temperatures the absorption of atoms does
 not really affect the structure of the model, but they do create a
 few strong absorption lines in the ultraviolet and visible part of
 the spectrum. The most prominent are the two CaII lines at 3968/3934
 \AA, the CaI line at 4227 \AA, the MgI triplet at 5167/5173/5184 \AA\ 
 the NaI doublet at 5890/5895 \AA, and the KI doublet at 7665/7699 \AA\ 
 \citep{walker14}. Still, it is the molecules that dominate the
 spectrum, completely obscuring most of the atomic lines except in the
 ultraviolet region. A more detailed look at the individual absorption of the molecules is 
 presented in Appendix \ref{s:detailed_spectrum}.

\begin{figure}
\centering
\includegraphics[width=\hsize]{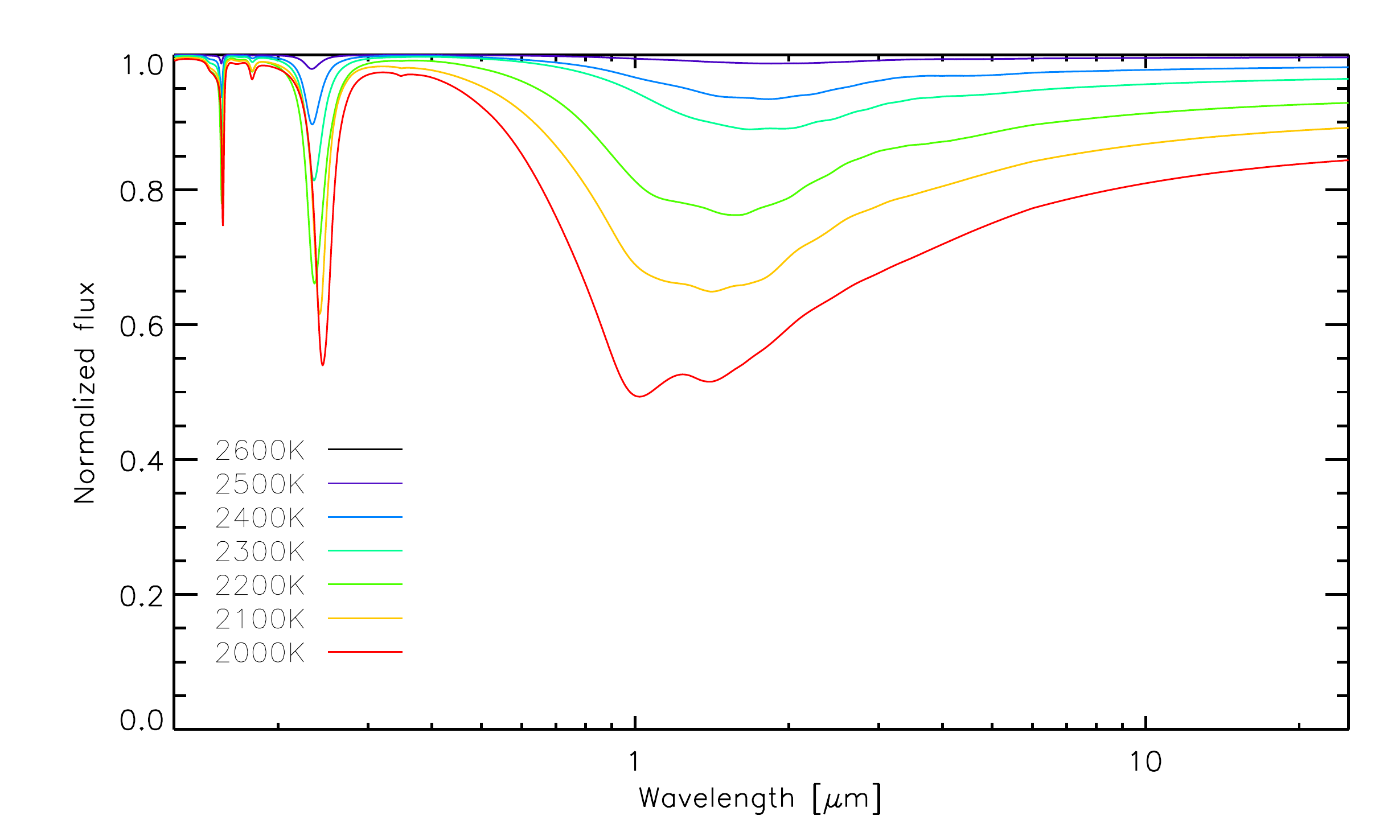}
\caption{{\color{black}Spectral contribution of dust (normalized with respect to the continuum flux) }
for models of decreasing effective temperature.}
\label{fig:dust-extinction}
\end{figure}

{\color{black}Figure \ref{fig:dust-extinction} shows how the increasing dust opacity 
affects the spectrum for models of decreasing effective 
temperatures. The normalized flux (the flux divided by the 
continuum flux) includes only the effect of dust on the spectrum.}
The dust opacity increases in a broad band 
that covers the optical and near-infrared wavelength regions, 
peaking at around 1-3 $\mu$m. For our coolest model, about half of 
the light is being blocked by the cloud layers at $\lambda \approx 1\ \mu$m. 
This is similar to the effect of water vapour in our cloud-free $T_\text{eff} = 2500$ K model (Figure \ref{fig:specfree}), and to (water) clouds in Earth's atmosphere. 

\begin{figure}
\centering
\includegraphics[width=\hsize]{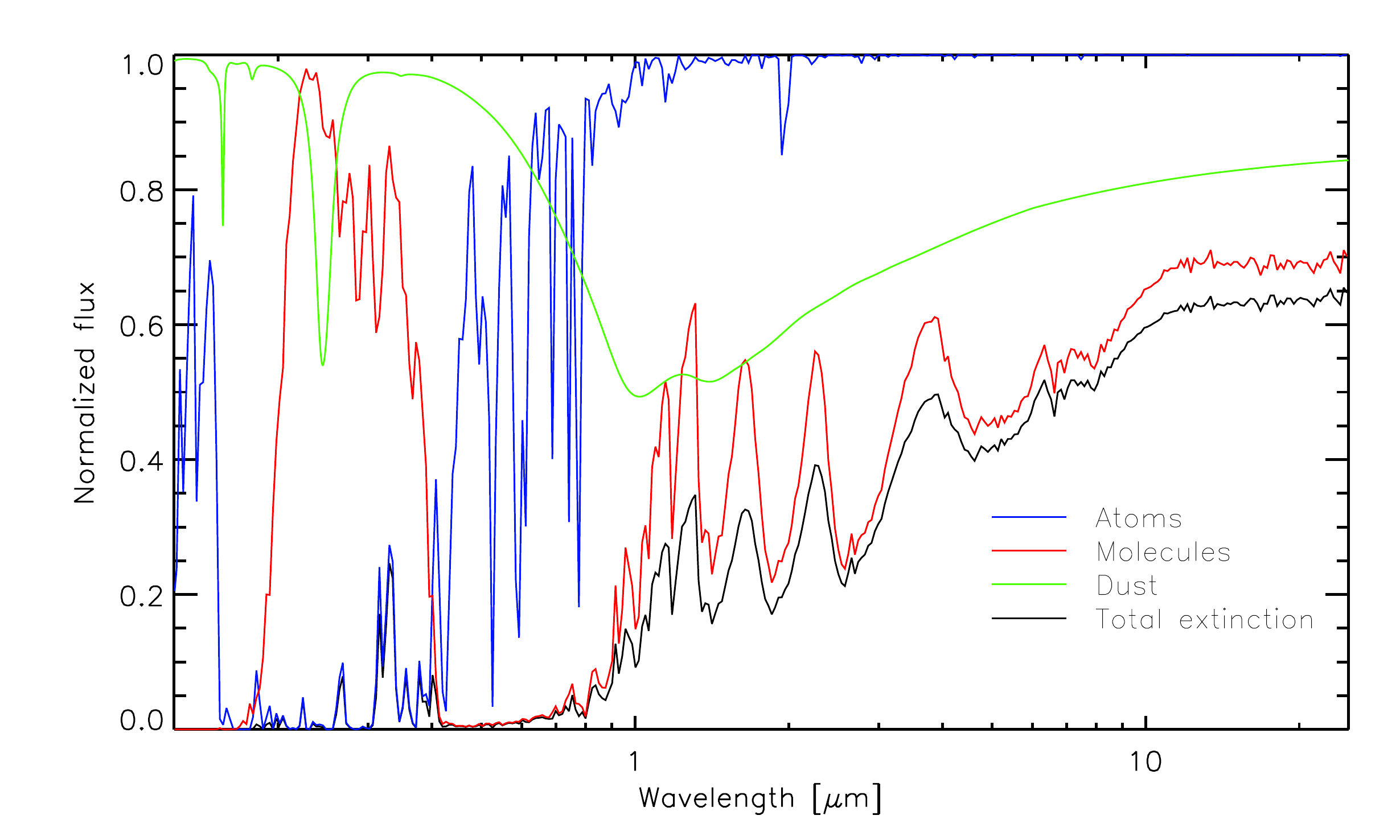}
\caption{The individual and combined effects of atomic opacity, molecular opacity 
and cloud opacity on the normalized flux (total flux divided by the continuum flux) 
of a cloud-forming {\sc Drift-Marcs}-model atmosphere with $T_\text{eff} = 2000$ K, 
$\log(g) = 4.5$ and [M/H] = 0.0.}
\label{fig:total-extinction}
\end{figure}

In Figure \ref{fig:total-extinction} we plot the effect of dust 
opacity (green) on the spectrum as a function of wavelength 
in comparison to the various gas opacity contributions 
(atoms -- blue, molecules --red) for a cloud-forming 
{\sc Drift-Marcs}-model atmosphere ($T_\text{eff} = 2000$ K, 
$\log(g) = 4.5$ and [M/H] = $0.0$). The whole wavelength range 
of {\sc Marcs} is shown. The dust extinction is most prominent 
in a broad band that covers the optical and near-infrared regions and peaks 
at around $\lambda = 1-2\ \mu$m, where it is comparable to or 
even greater than the molecular absorption. For $\lambda > 10\ \mu$m 
the dust extinction has a noticeable dampening effect on the 
molecular absorption bands, which would otherwise have completely 
dominated the spectrum. The two sharp peaks at short wavelengths are
numerical artifacts pointing to challenges with the Mie calculations.
They, however, occur in the ultraviolet part of the spectrum where the
opacity is heavily dominated by atomic absorption and therefore does
not have an effect on the spectrum (nor on the structure of the model). 
The dust extinction increases with
decreasing effective temperatures. 

\subsection{Optical depth}
\begin{figure}
\hspace*{-0.7cm}
\includegraphics[width=10cm]{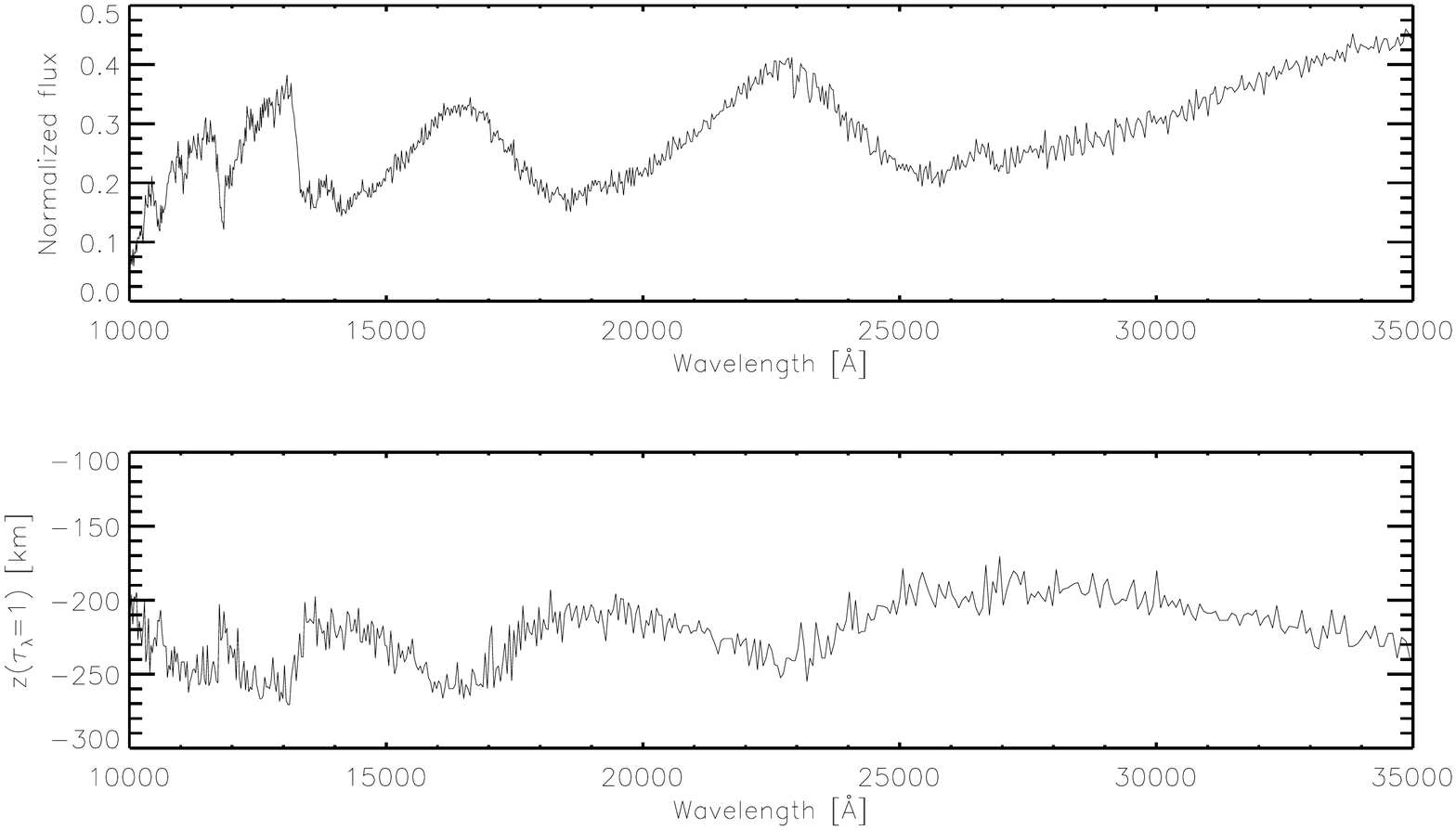}
{\ }\\*[-1.7cm]
\caption{The total nominal flux (top) and the atmospheric depth,
  $z(\tau_\lambda = 1)$ for cloud-forming {\sc
    Drift-Marcs}-model atmosphere for $T_{\rm eff}$=2000 K, $\log(g)=4.5$, 
    [M/H]=0.0. {\color{black}The total geometric extension of the atmosphere from 
    $\log(\tau_\text{ROSS}) = 2$ to $\log(\tau_\text{ROSS}) = -10$ is 333 km.}}
\label{fig:spectau}
\end{figure}

The opacity of the atmosphere changes with wavelength and therefore 
so does the optical depth $\tau(\lambda)$. We can determine the optical 
depth of the atmosphere from our synthetic spectrum and thereby estimate 
how deep into the atmosphere we can see at a specific wavelength. 
In Figure~\ref{fig:spectau} we plot the total normalized flux (top) and the 
geometrical depth, $z(\lambda)$, where $\tau(\lambda)$= 1 in a cloud-forming 
atmosphere with $T_\text{eff} = 2000$ K, $\log(g) = 4.5$ and [M/H] = 0.0. 
When the opacity is high, the flux is low and we cannot see as far into the atmosphere
as when the opacity is low and the flux is high. Figure~\ref{fig:spectau} 
demonstrates that the overall flux in the near infrared is absorbed by $>50\%$. An additional
flux variation of $\approx10\%$ translates into a $\Delta z(\lambda)\approx 50$ km
which is 15\% of the total geometrical extension of the atmosphere of a $\log(g)=4.5$-type
ultra cool object as seen in Figure~\ref{fig:spectau}. This is the cause of the observable variation in an exoplanet 
transit depth as a function of wavelength and is the direct cause that 
transit observations can be translated into exoplanetary spectra and structure. 
%(see Table \ref{tab:flux} and Figure \ref{fig:wasp19b}). 

\subsection{Comparison to observed spectra}
Ultra cool dwarfs emit the majority of their radiation flux at near-infrared 
(NIR) wavelengths and their discovery and classification is therefore 
mainly conducted by NIR spectroscopic instruments. One such 
instrument is the SpeX spectrograph mounted on the 3 m NASA 
Infrared Telescope Facility, which provides moderate and low resolution 
broad-band NIR spectra \citep{rayner03}. SpeX spectra has proved 
ideal for NIR classification, characterization of atmospheric and 
physical properties as well as testing atmosphere models 
\citep{burgasser14}, 
%We have selected a few hundred SpeX spectra of 
%M-dwarfs (159) and L-dwarfs (261) from the online SpeX Prism Spectral 
and is made readily available from the online SpeX Prism Spectral
Libraries\footnote{http://www.browndwarfs.org/spexprism} that we have compared 
with our synthetic {\sc Drfit-Marcs} spectra.

The SpeX spectra are all normalized, have a resolution of $R = 
\lambda/\Delta\lambda \approx 120$, and span a wavelength range of $\lambda \approx 
0.65 - 2.55\ \mu$m. {\color{black}For a single comparison, we re-sampled our synthetic 
spectrum to match the resolution and range of the observed spectrum and 
then fitted the synthetic spectrum to the observed spectrum by simply scaling 
the synthetic spectrum. We used the non-linear least squares curve fitting 
routine {\sc MPFIT} \citep{markwardt09} which identifies the best fit as the 
one with the lowest value of 
\begin{equation}
\chi^2 = \left( \sum_{i=1}^N \frac{(A \cdot f_\text{synth,i} - f_\text{obs,i})^2}{\sigma_\text{obs,i}} \right) / (N-1),
\end{equation}
where $A$ is the scaling factor, the only free parameter, and $N$ is the number of data points. We repeated 
this process for every combination of synthetic and observed spectrum, 
in the end identifying the best fitting synthetic spectrum for each 
observed spectrum as the one with the lowest value of $\chi^2$.}

{\color{black} Most of the $\chi^2$ values were in the range of $\chi^2 \approx 1.5-15$ 
with a few very large exceptions. Since our grid serves as an indication of the direction 
we are going in with our models, we expect that a good deal of the fits will be considerably 
improved once we have computed a more complete grid that includes variations in surface 
gravity or metallicity. We are therefore wary of systematic offsets in our fits and only 
consider the best fit of a few selected spectral sub classes, 
where $\chi^2 < 2.5$ is low enough to assume a true match between 
synthetic and observed spectra.}

Representative stars within the parameters of our small grid for the present project and their best fit 
models are presented in Table \ref{table:fitted-spectra}
and Figs.~\ref{fig:specMdwarfs} and \ref{fig:specLdwarfs}. These objects have not been 
presented in \cite{witte2011}. Here, we focus on mid- to late-type M-dwarfs (Section~\ref{ssM}),
early- to mid-type L-dwarfs (Section~\ref{ssL}), and on warm, giant gas
planets (Section~\ref{ssGP}). We specifically address the giant gas
planet WASP19b as one example.

\begin{table*}
\caption{The name, spectral type and data reference for the observed 
spectra together with the parameters of our best fit model. All models have 
$\log(g) = 4.5$ and $[M/H]=0$.}      
\label{table:fitted-spectra}      
\centering          
\begin{tabular}{l l r | c c}
\hline\hline       
\multicolumn{3}{l}{{\textbf{Object}}} 							& \multicolumn{2}{| l}{{\textbf{Best fit model}}} \\
Name 			 		& SpT 	& Reference			& $T_\text{eff}$& $\chi^2$ \\
\hline
2MASS J12471472-0525130 	& M4.5	& \citet{kirkpatrick10}		& 3000 K		& 1.41	 \\
Gliese 866AB				& M5.6	& \citet{burgasser08}		& 2900 K		& 1.60	 \\
2MASS J11150577+2520467	& M6.5	& \citet{burgasser04}	& 2800 K		& 2.02	 \\
VB 8 					& M7		& \citet{burgasser08}	& 2800 K		& 1.77	 \\
2MASS J17364839+0220426 	& M8		& \citet{burgasser04}	& 2700 K		& 2.05	 \\
2MASS J11240487+380854	& M8.5	& \citet{burgasser04}	& 2600 K		& 2.49	 \\
2MASSW J0320284-044636	& M8/L0.5& \citet{burgasser08}	& 2500 K		& 2.22	 \\
2MASS J15500845+1455180	& L2		& \citet{burgasser09} 	& 2000 K		& 2.33	 \\
2MASSW J0036159+182110	& L3.5	& \citet{burgasser08}	& 2100 K		& 2.09	 \\
2MASS J1104012+195921	& L4   	& \citet{burgasser04}	& 2000 K		& 2.08	 \\
SDSS J154849.02+172235.4	& L5  	& \citet{chiu06}		& 2000 K		& 1.91	 \\
2MASS J14162409+1348267	& L6		& \citet{schmidt10}		& 2300 K		& 2.20	 \\
\hline
\end{tabular}
\end{table*}

\subsubsection{Mid- to late type M dwarfs}\label{ssM}
\begin{figure}
\hspace*{-0.7cm}
\includegraphics[width=10cm]{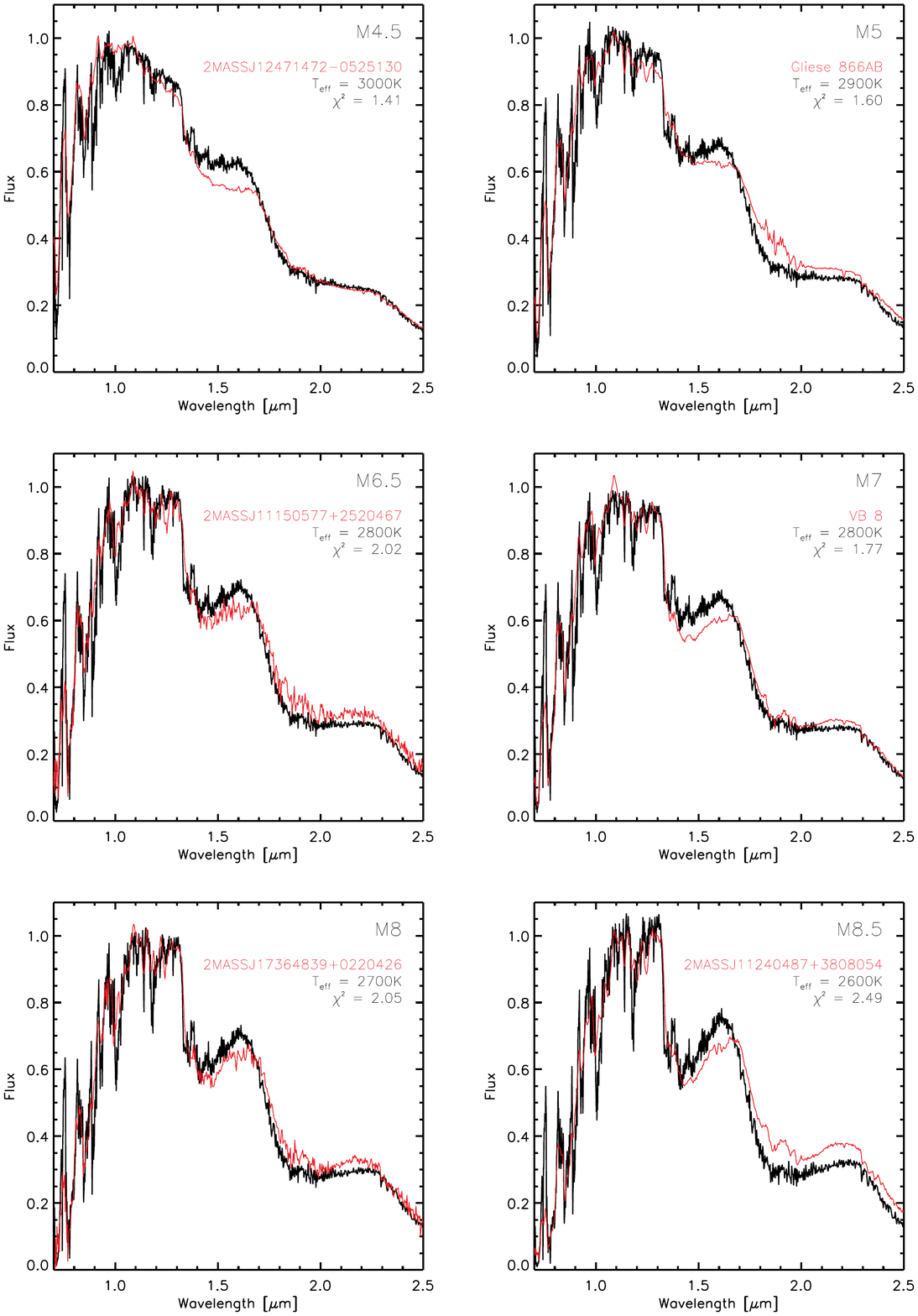}
{\ }\\*[-1.0cm]
\caption{Mid- and late M-dwarf SpeX observations  fitted with {\sc Drift-Marcs}.}
\label{fig:specMdwarfs}
\end{figure}

In Figure~\ref{fig:specMdwarfs} we present the comparison between 
the observed spectra of six M-dwarfs and our best fit models. The 
earliest subtype that can be fitted by our models is M4.5. With an 
effective temperature of $T_{\rm eff}$ = 3000 K its atmosphere is dust free, 
and its spectrum is generally well modeled by the synthetic spectrum of our 
model. The famous TiO bands of M-dwarfs dominate the
total absorption from $0.7-1.0\ \mu$m, only disturbed slightly
by VO at 0.8 $\mu$m and CrH at $0.85-0.9\ \mu$m. The absorption
band in the model at $\lambda=1.0\ \mu$m is a mix of CrH, TiO and FeH in
order of influence, but it is not observed in the spectra of this star.
For $\lambda>1.3\ \mu$m the broad absorption bands of H$_2$O become
the main absorption features and they stay almost undisturbed by other
molecules and atoms except at $\lambda= 2.3-2.4\ \mu$m where CO
absorption causes the small fluctuations.{\color{black} The absorption is
somewhat underestimated at $\lambda= 1.4 -1.7\ \mu$m for most of 
the models and slightly overestimated at $\lambda=1.8-2.3\ \mu$m. 
We note however that the deviation is not correlated with $T_{\rm eff}$, 
and that the magnitude of the two deviations are not correlated with 
one another. We therefore conclude that the mismatch most likely is 
due to chemical abundance effects beyond the range of our present 
grid.}

None of the M-dwarfs reach effective temperatures below 
$T_\text{eff} = 2600$ K, and if any cloud formation takes place in 
their atmosphere, it plays no significant role in their spectra. As their 
effective temperatures decrease, the peak of their spectrum shifts 
towards longer wavelengths. The intensity of the TiO bands grows larger
and are blended with the increasing absorption of VO and CrH. The
absorption of CaH also increases at $\lambda= 0.7-0.75\ \mu$m
but has a very small effect on the spectrum. The increase in absorption of CrH, VO
and FeH at $\lambda= 1.0\ \mu$m is well matched by the
models. Finally, the absorption of H$_2$O in the infrared increases
significantly as well, each band growing deeper with decreasing
effective temperature. With the massive suppression of the continuum due to
the absorption of molecules, the coolest M-dwarfs are clearly far away
from being ideal black body radiators. As demonstrated in
Figure~\ref{fig:temperature-pressure2},
clouds barely affect the atmosphere of
objects with $T_{\rm eff}>2600$ K.

\subsubsection{Early- and mid type L-dwarfs}\label{ssL}
\begin{figure}
\hspace*{-0.7cm}
\includegraphics[width=10cm]{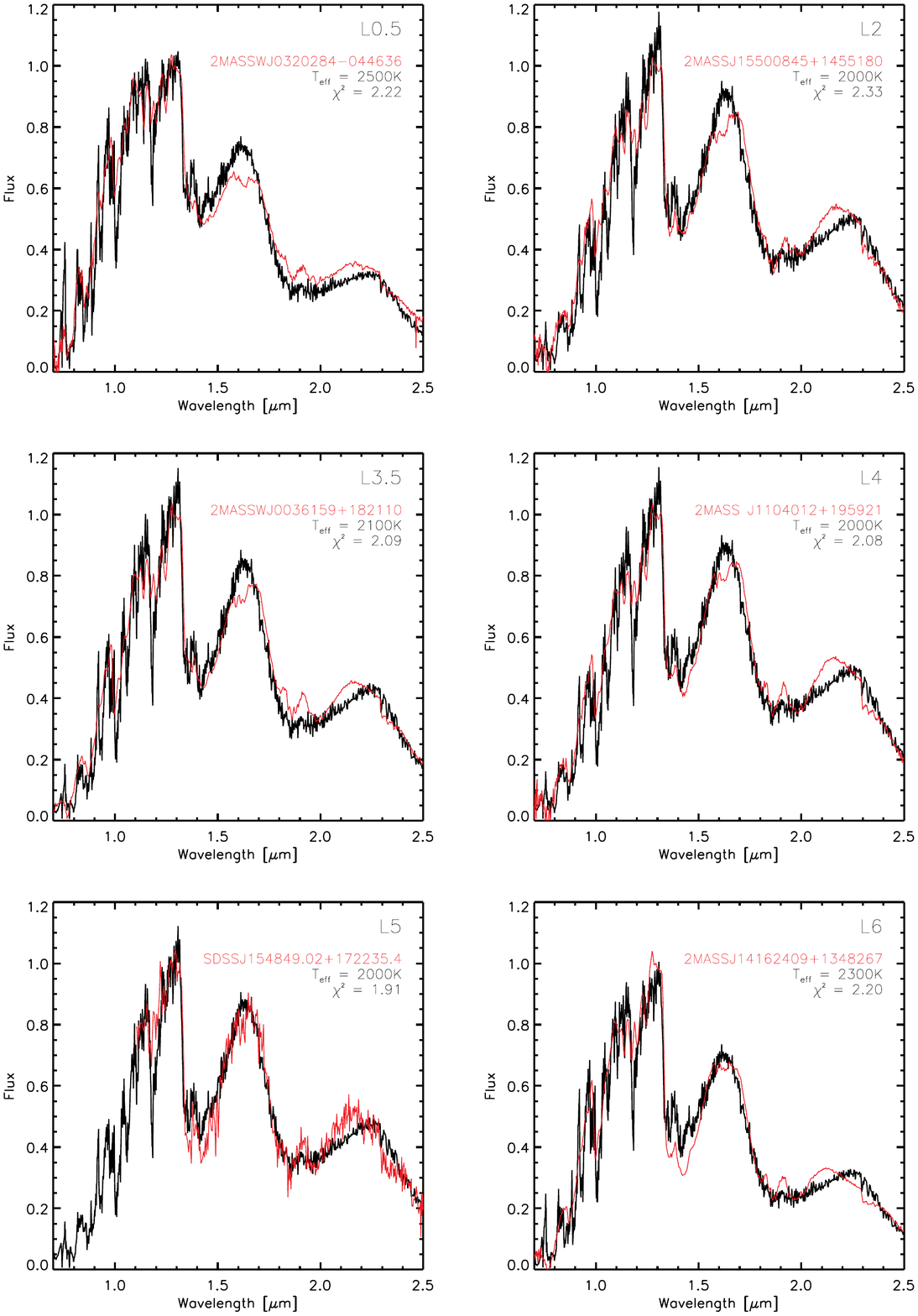}
{\ }\\*[-1.0cm]
\caption{Early- and late L-dwarf SpeX observations  fitted with {\sc Drift-Marcs}.}
\label{fig:specLdwarfs}
\end{figure}

In Figure~\ref{fig:specLdwarfs} we present the comparison between 
the observed spectra of six L-dwarfs and our best fit models. The latest 
sub type that can successfully be fitted by the model grid in our present work
is L6. For later sub types $\chi^2$ becomes too large as the effective 
temperatures go below our lowest grid temperature of
$T_\text{eff} = 2000$ K. For decreasing effective temperatures, the 
absorption from $0.7-1.0\ \mu$m gradually becomes characterized
by equally strong TiO and VO bands. The absorption of CaH and CrH also
becomes stronger in that region, but since their bands tend to
coincide with the stronger TiO and VO bands, they do not affect the
spectrum that much. The absorption feature at $\lambda = 1.0\ \mu$m is the
result of a peak in CrH absorption as well as absorption from TiO and
FeH. The other noticeable absorption features at $\lambda = 1.2\,\mu$m is caused by the
superposition of the absorption peaks of CrH, H$_2$O, VO and CaH and
FeH. 

The infrared part of the SpecX spectral region is dominated by three strong absorption 
features at 1.4 $\mu$m, 1.9 $\mu$m and 2.5 $\mu$m, and corresponding opacity minima at 
1.6 $\mu$m and 2.2 $\mu$m. The intensity of the absorption bands are determined by a 
temperature, pressure and elemental abundance dependent combination of CO and H$_2$O 
and could also include contributions by other species with yet incomplete opacity data. The 
flux at the intensity minima are to a large extend determined by the more continuum-like dust 
absorption. The slight mismatch between our synthetic spectra and the SpecX infrared 
observations could therefore be due to incomplete inclusion of a combination of any of these 
factors, but it is probably more likely (since there is no clear $T_{\rm eff}$ dependence on the 
quality of the fits) to be due to the smallness of the grid of models yet, which does not allow 
us to vary the chemical abundances and gravity sufficiently for more accurate fits to the 
observed spectra. The present paper is, however, not aiming either at determining the 
parameters of the presented objects by detailed spectral fitting, but rather to develop the 
basic principles of incorporating self-consistent dust formation into the gas phase atmospheric 
computation, and test whether such models relate realistically to observations. 
As such Figures \ref{fig:specMdwarfs} and \ref{fig:specLdwarfs} fully serve 
their purpose of demonstrating that this has been achieved, and we will leave the 
detailed matching to a future paper with a more extended grid.
%In the infrared, where the absorption of cloud particles becomes more and
%more significant, all the observed L-dwarf spectra show a growing
%bump at  $\lambda = 1.9\,\mu$m that our models can not reproduce. Since it is
%located right where the cloud particle extinction peaks (see Figure \ref{fig:dust-extinction}), 
%a possible explanation could be that our models produce a little too much cloud opacity. 

In general we see that the optical region is dominated by TiO and VO absorption 
bands, the near-infrared region by strong metal hydride bands (CrH, FeH, and CaH), and 
the infrared region by the broad, cloud-opacity dampened H$_2$O absorption bands.

{\color{black}We note that our best fit model to 2MASS J1416 is several hundred Kelvin 
warmer than a typical L6 type dwarf. This particular L dwarf has been identified as an 
unusually blue object for its spectral type \citep{bowler10}, and it is therefore likely that 
non-solar metallicities or other effects makes it impossible for our small grid of models 
to fit it correctly. 

Furthermore, 2MASSW J0320 might be an unresolved late M + T dwarf binary system 
\citep{burgasser08} and can in that case not be fitted well by a single model spectrum.}

\subsubsection{WASP-19b}\label{ssGP}
Hot Jupiters have deep hydrogen-helium atmospheres. Some of them orbit
so close to their parent stars that they have surface temperatures larger 
than $T = 2000$ K. We apply {\sc Drift-Marcs} to model such an atmosphere where
we do not yet take into account the irradiation by the host star. 

The atmosphere of WASP-19b was modeled by \cite{anderson13} using the 
spectral retrieval methods
developed in \cite{madhusudhan09,madhusudhan10,madhusudhan11}, 
which utilize parametric ($T_{\rm gas}$, P$_{\rm gas}$) structures 
in combination with a cloud-free gas made of H$_2$, H$_2$O, CO, CH$_4$, CO$_2$, 
and NH$_3$. In the retrieval approach, the ($T_{\rm gas}$, P$_{\rm gas}$) structure and 
the molecular abundances are fitting parameters used to retrieve the observed spectrum.
In the {\sc Drift-Marcs} approach, on the other hand, the ($T_{\rm gas}$, P$_{\rm gas}$) structure and 
the abundances of the individual gas- and dust-species are computed
from first principles self-consistently with the radiative transfer, energy balance, opacities, and 
dust formation, as described above. There are therefore no free parameters in {\sc Drift-Marcs} spectrum
simulations (but, as mentioned above, irradiation is not yet included in the version 
presented here), and the best fit model gives direct information about the temperature profile 
and the chemical composition of the planetary atmosphere. 

%The radiative transfer was solved, and the retrieval approach applied to
%determine the parameters that determine the (T$_{\rm gas}$, P$_{\rm gas}$) structure 
%and the abundance of the gas opacity carrier. Here, on the other hand, 
%we model the observations with fully self-consistent model computations.

WASP-19b is a transiting exoplanet with a mass of M$_{\rm p} =
1.165$M$_{\rm J}$ and a radius of R$_{\rm p} = 1.383$R$_{\rm J}$ in a
close orbit around its parent star with a period of only $P = 0.789$
days, as determined from transit and radial velocity measurements. 
It is therefore classified as a hot Jupiter. The day-side flux
of WASP-19b has been measured by observing the occultation of the
planet by its parent star with the Spitzer Space Telescope. The
relative flux of the planet with respect to its star is presented in
Table~\ref{tab:flux} (Table 4, \cite{anderson13}).  

\begin{table}
\caption{The relative flux of the exoplanet WASP-19b with respect to its star at different wavelengths (Table 4, \citet{anderson13}).}      
\label{tab:flux}      
\centering          
\begin{tabular}{l l l}
\hline\hline       
Wavelength	 	& $F_\text{p}/F_\star$ 		& Reference \\
\hline     
$1.6\ \mu \text{m}$	& $0.00276 \pm 0.00044$		& \citet{anderson10} \\
$2.09\ \mu \text{m}$	& $0.00366  \pm 0.00067$	& \citet{gibson10} \\
$3.6\ \mu \text{m}$	& $0.00483 \pm 0.00025$		& \citet{anderson13} \\
$4.5\ \mu \text{m}$	& $0.00572 \pm 0.00030$		& \citet{anderson13} \\
$5.8\ \mu \text{m}$	& $0.0065 \pm 0.0011$		& \citet{anderson13} \\
$8.0\ \mu \text{m}$	& $0.0073 \pm 0.0012$		& \citet{anderson13} \\
\hline
\end{tabular}
\end{table}
{\color{black}The parent star of WASP-19b is a G8V type
star given by \citet{anderson13} as $T_{\rm eff}$ = 5475 K, $\log(g)$ = 4.43 and [M/H] = 0.02. We
used {\sc Drift-Marcs} to compute a stellar model atmosphere with these 
parameters and calculated its synthetic spectrum, the flux $f_\star$. 
The relative flux $F_\text{p}/F_\star$ which we receive on the Earth from the two is
\begin{equation}
%\frac{F_\text{p}}{F_\star} =  \frac{f_\text{p} \cdot \pi R_\text{p}^2}{f_\star \cdot \pi R_\star^2} = \frac{f_\text{p}}{f_\star} \left( \frac{R_\text{p}}{R_\star}\right)^2,
\frac{F_\text{p}}{F_\star}  = \frac{f_\text{p}}{f_\star} \left( \frac{R_\text{p}}{R_\star}\right)^2,
\end{equation}
where $(R_\text{p}/R_\star)^2 = 0.02050 \pm 0.00024$ is the planet-to-star area ratio. 
We calculated $F_\text{p}/F_\star$ for each of our cloud forming 
{\sc Drift-Marcs} models, setting $f_\text{p}$ as their respective fluxes.}
Comparing these synthetic planet-to-star fluxes with the observed planet-to-star flux, 
we found that it  is best described by our cloud forming, non-irradiated model with 
$T_{\rm eff}$ = 2600 K, $\log(g)$ = {\color{black}3.2} and [M/H] = 0. 

\begin{figure}
\hspace*{-0.7cm}
\includegraphics[width=10cm]{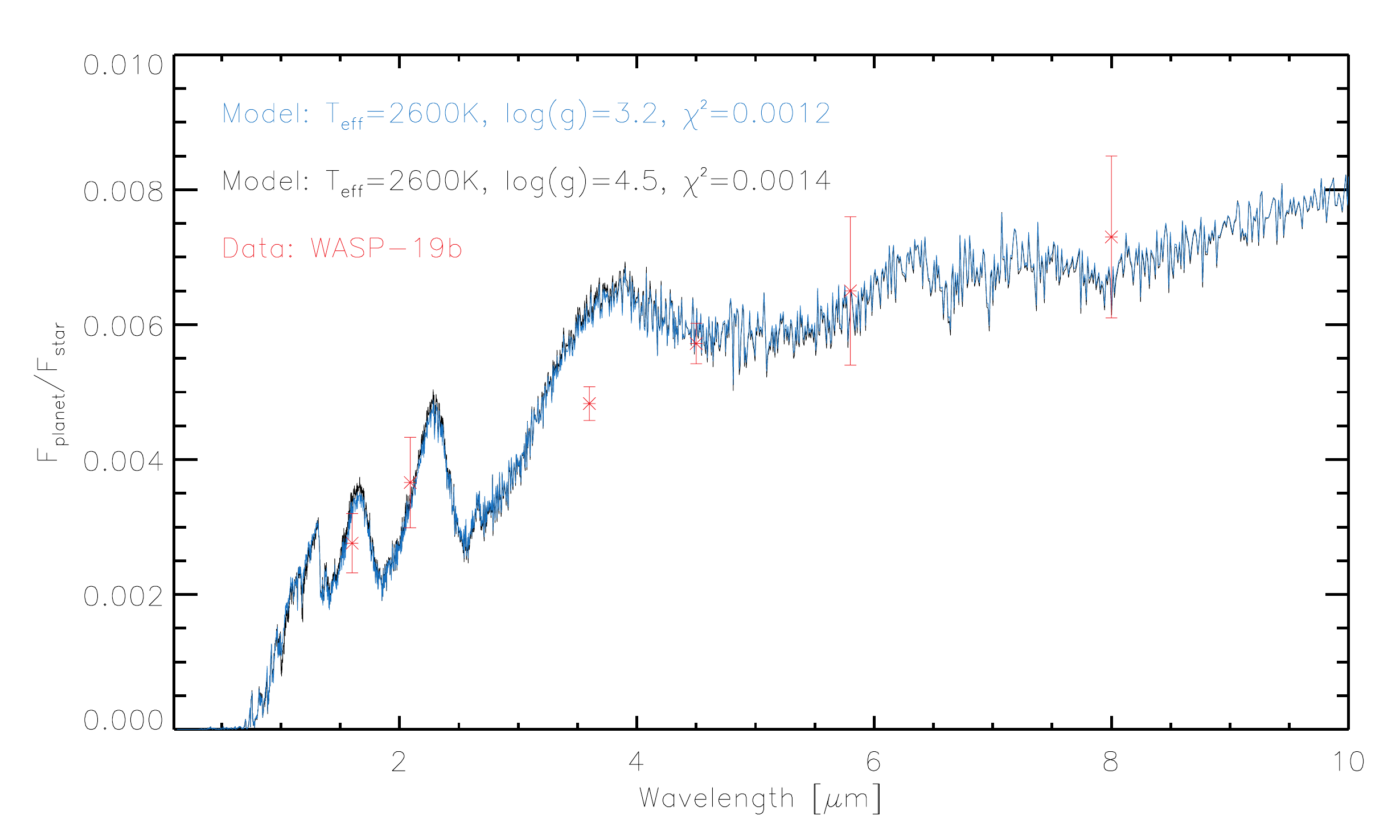}
{\ }\\*[-1.0cm]
\caption{The best-fit synthetic transit spectrum for WASP-19b for 
$\log(g) = 4.5$ (blue) and $\log(g) = 3.2$ (green), 
  based on {\sc Drift-Marcs} model atmosphere simulations for the star
  ($T_{\rm eff}$ = 5475 K, $\log(g)$ = 4.43, [Fe/H] = 0.02) and for the
  planet. The Spitzer data points are plotted in red. We derive that the 
  temperature of the planet is $T_{\rm eff}$ = 2600 K.}
\label{fig:wasp19b}
\end{figure}

{\color{black}The spectrum is quite insensitive to the value of log($g$), which therefore 
cannot be determined from the spectrum (log($g$) = 3.2 and 4.5 gives 
basically identical IR spectra at this resolution as is demonstrated in 
Figure~\ref{fig:wasp19b}). Instead we adopt the value log($g$)=3.2 from 
determination of the planetary mass and radius, as given in \citet{anderson13}. 
Our preliminary grid presented here is only calculated for solar metallicity (in agreement 
with what is found for WASP-19, but WASP-19b could obviously be metallicity enhanced). 
Our model fit is therefore nothing other than a rough first temperature fit and, for the purpose 
of the present paper, mainly a working demonstration that our model method is able to 
self-consistently reproduce even hot exoplanetary spectra. There are no other self-consistent 
temperature estimates we could compare to in the literature, but \citet{anderson13} attempted 
an estimate of the planetary effective temperature by calculating a brightness temperature 
obtained by dividing a Planck function for various guesses of the effective temperature of the 
planet with a Planck function of the stellar effective temperature, and fitting this ratio to the 
Spitzer measurements. In this way they found values between 2260\, and 2750\,K (depending on 
the filter they fitted to). They also estimated the planetary equilibrium temperature based on the 
known effective stellar temperature and the planetary orbital size, and by assuming a planetary 
albedo of zero. In this way they reached effective temperatures of WASP-19b between 2040\,K 
and 2614\,K (or actually, between 2433\,K and 3109\,K when correcting for a missing $\sqrt{2}$ 
in their formula for calculating $T_{\rm eff}$ as given in the caption to their Table\,3). The range 
in temperature reflects a range in assumed efficiency in energy transport from day- to night-side 
of the planet. They also analysed whether a temperature inversion was visible in the measured 
flux distribution, and concluded that their temperature inversion profile was inconsistent with the 
observed flux distribution (however, seemingly with lacking the spectral features of the inversion 
in their computed spectra, due to lacking chemical equilibrium and relevant opacities in the computations). 
Our estimate of $T_{\rm eff}$ = 2600\,K is in good agreement with the estimate by Anderson et al of an 
atmosphere with solar C/O ratio, albedo zero and no temperature inversion. This is quite encouraging, 
because it qualitatively points at an atmosphere with low albedo, that is relatively clear, absorb most of 
the incoming energy in the bottom of the atmosphere (with winds that will transport energy to the backside, 
but not so efficient that the planet reach equal day and night temperature), and has no sign of a strong 
temperature inversion (with the caution that neither we nor Anderson et al really have analysed the effect 
of a temperature inversion, due to the two different computational limitations mentioned above). 
Figure~\ref{fig:wasp19b} shows the comparison of our synthetic spectrum and the Spitzer observations.}
%Figure~\ref{fig:wasp19b} shows
%that our results fit  the observations reasonably well without the need
%of a temperature inversion. \cite{anderson13} derived $T_{\rm eq}\approx
%2000-2600$ K(\footnote{$T_{\rm eq}=f^{0.25}T_{\rm
%eff}\sqrt{R_*/(2a)}$ with $f$ the redistribution parameter, $R_*$
%the stellar radius, $a$ the semimajor axis}) (depending on
%heat-redistribution parameter). For instant re-radiation the heat distribution parameter 
%is $f = 8/3$ and $T_\text{eq} \approx 2614$ K. This number is in fine 
%agreement with our result, but a big advantage of our calculation is 
%that it is self-consistent and hence predict the physical and chemical 
%structure of the atmosphere, at the same time as reproducing the 
%spectrum.

%\begin{figure}
%\hspace*{-0.7cm}
%\includegraphics[width=10cm]{images/Fig741.pdf}
%{\ }\\*[-1.0cm]
%\caption{Testing different H$_2$O line lists in a cloud-free {\sc
%    Marcs} model atmosphere for T$_{\rm reff}$ =2900K, log(g) = 4.5
%  and [M/H] = 0.0 in comparison to an SpeX observation (black) for the
%  M-dwarf 2MASS\,J0058-1747. {\bf 1st panel:} Synthetic {\sc Marcs}
%  spectra using different lines lists (SCAN -- blue, HITEMP -- red,
%  Partridge \& Schwenke -- green. {\bf 2nd -- 4th panel:} Comparsion
%  of synthetic spectra with different H$_2$O line lists to
%  observation.}
%\label{fig:H2O}
%\end{figure}

%-----------------------------------------------------------------------------------------------------------
% DISCUSSION
%-----------------------------------------------------------------------------------------------------------
\section{Cloud particle porosity}\label{s:disc}
Material properties are an essential input for every model. The
challenge of obtaining such input has recently been outlined in
\cite{fort2016}. Here we shortly discuss the effect of porosity on the 
cloud opacity.

\begin{figure}
\hspace*{-0.7cm}
\includegraphics[width=10cm]{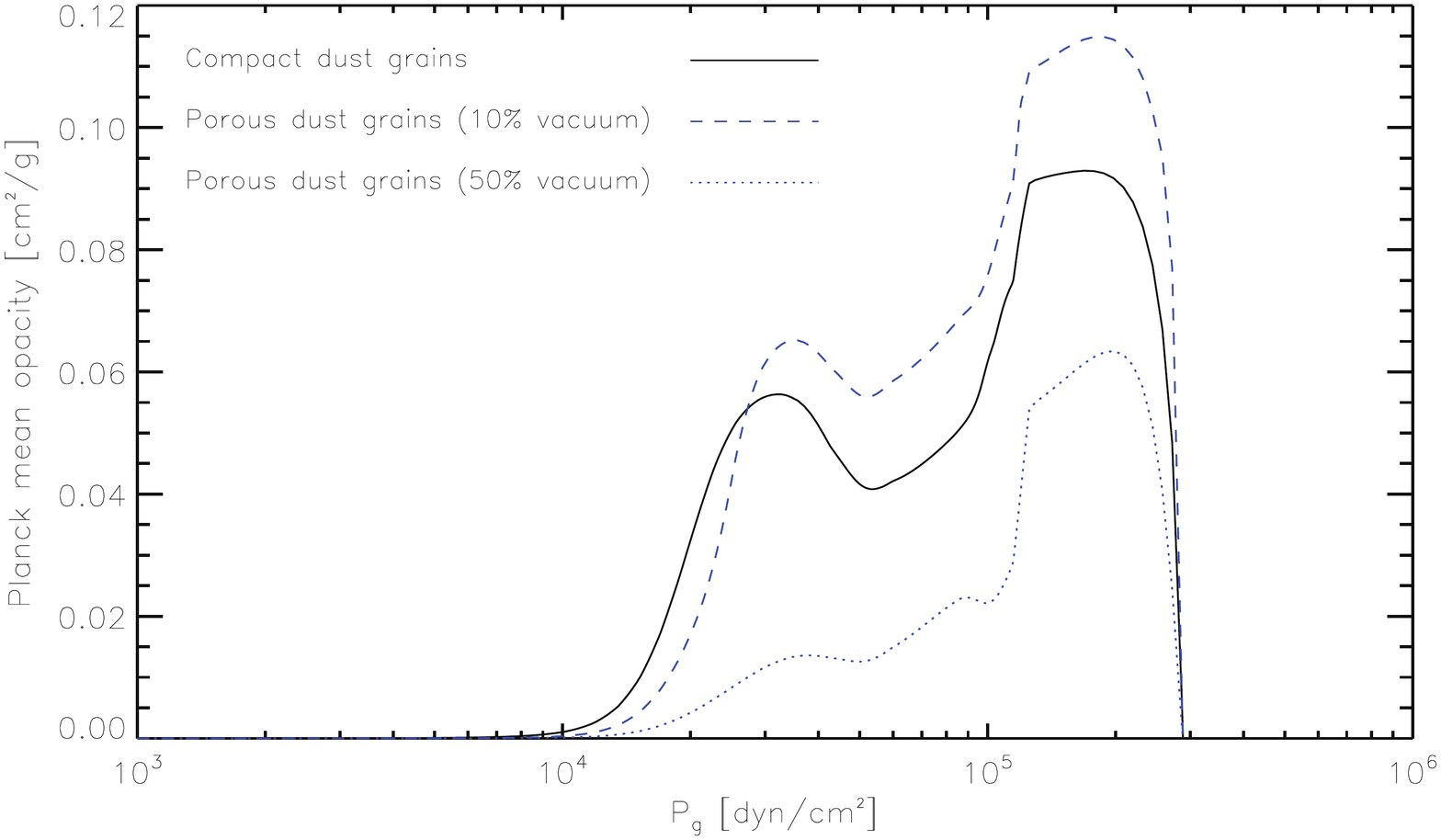}
{\ }\\*[-1.5cm]
\caption{Testing the effect of porosity on the cloud opacity ($T_{\rm eff} = 2000$ K, $\log(g)$ = 4.5
  and [M/H] = 0.0). The plotted cloud opacity is the Planck mean opacity.}
\label{fig:por}
\end{figure}

Figure~\ref{fig:por} shows that the porosity of cloud particles can
have a considerable effect on the opacity. We chose to represent the
effect in terms of local Planck mean opacities as this allows us to 
plot a meaningful measure of the opacity as a function of the whole 
atmospheric extension. We compare the integrated
opacity for one example cloud-forming model atmosphere ($T_{\rm eff}$
= 2000 K, $\log(g)$ = 4.5, [M/H]=0.0) for three types of cloud particles:
one is compact (using the results directly from
Figure~\ref{fig:dust-info},~\ref{fig:dust-volume}; solid line), one
contains 10\% vacuum (dashed line) and one contains 50\% vacuum
(dotted line). {\color{black}We test this by adding the vacuum as an eighth condensate 
and then scaling the contribution of each condensate such 
that the total volume of a single dust grain - and thereby also its surface 
area - remains the same, while its mass decreases for increasing porosity}.
The model atmosphere has not
been iterated with the new cloud opacity, we have just re-calculated
the opacity of the cloud layer in the already converged model to
illustrate if there is an effect. Interestingly, we see that by
increasing the porosity slightly the cloud grains become more
opaque. If the porosity is too high the opacity drops again since the
light can pass unhindered through a large part of the cloud grains.

Porosity could arise if the cloud
particles do not attain a compact shape during their formation or
evolution, but rather develop fractal shapes instead. We are familiar 
with this process from Earth's atmosphere as "snow". Comets are examples 
of a different kind of porosity. It is not clear whether potential porosity 
can sustain as the cloud particles fall into deeper
atmospheric layers where their frictional interaction with the gas
increases, which then would lead to a compactification or break-off of
dangling structures. A more realistic scenario for relatively hot atmosphere 
could be that different materials evaporate at different temperatures, while others
remain thermally stable throughout the entire atmosphere. 

%-----------------------------------------------------------------------------------------------------------
% CONCLUSIONS
%-----------------------------------------------------------------------------------------------------------
\section{Conclusions}

The coming years and decades will see a substantial technological development that
will make it possible to obtain direct spectra of increasing quality of nearby exoplanets.
Reliable interpretation of such high quality spectra will require detailed complex 
self-consistent model atmospheres, which at the same time will make it possible to 
reliably quantify such exciting features as potential biomarkers in the atmosphere, and 
hence open a route for the first scientific discussions of possible life forms on nearby
extrasolar planets. With these long-term goals in mind we have taken the first steps
to combine two well tested computer codes from stellar atmospheric theory, namely the 
{\sc Marcs} radiative and convective equilibrium code for gaseous atmospheres and the {\sc Drift}
dust and cloud formation code. In combination, the {\sc Drift-Marcs} code that we here have 
presented for the first time, is able to compute 
self-consistent model atmospheres that can include both
radiative-convective energy transport, the chemical equilibrium between
both gas and dust species, as well as cloud formation and cloud destruction.   
These are necessary ingredients to compute self-consistent models of exoplanetary 
atmospheres, and the exercise serves a double purpose, namely to pave the way for 
{\sc Drift-Marcs} self-consistent general exoplanetary models and to increase the accuracy of 
the stellar models of the type of stars, M- L- and T-type stars (and brown dwarfs), 
whose orbiting exoplanets we already today are able
to obtain crude spectra of. 

M-, L-, and T-dwarfs are very attractive targets when searching for new 
exoplanets by indirect methods. Their relatively small mass and size provide stronger
signals for detection with the radial velocity, astrometry, and transit
methods. An inherent problem of these exoplanet search
methods is that the uncertainty of the properties of the host star propagates
to the properties of its planet. It is therefore crucial that the stellar models 
linking the observations of a star to its properties are as precise as possible,
and the ultra-cool dwarf stars are much more complex to model than their 
larger and hotter cousins, mainly because their temperatures are low enough for mineral
clouds to form in their atmospheres. We have demonstrated when and how the mineral
cloud formation starts to play a role for the atmospheric structure of our models, and
we have shown that emergent spectra based on our {\sc Drift-Marcs} model atmospheres are in
good agreement with observed spectra for the whole range of spectral types from 
mid-type M-dwarfs to late-type  L-dwarfs ($T_{\rm eff}$ = 3000\,K to 2000\,K). The
{\sc Drift-Marcs} code is therefore already in its present form a reliable tool to 
accurately determine the stellar parameters and hence improve the
parameters of exoplanets orbiting cool dwarf stars.

Hot Jupiter exoplanets orbiting solar-type and warmer stars are themselves of 
comparable ($T_{\rm eff}$, log($g$)) values to the ultra-cool dwarf stars, and
one would expect them to have slow or tidally locked rotations. They will therefore 
to a large extend resemble the ultra-cool dwarf stars, and can therefore to a first
approximation be modeled in the same way as these.
Crude spectra or photometric data points can already today be obtained for a few
hot Jupiter exoplanets by subtracting the stellar spectrum during occultation
from the stellar spectrum with the exoplanet in different phases 
(i.e. positions of its orbit). We therefore tested our computed {\sc Drift-Marcs} 
synthetic spectra against photometric data of the hot Jupiter WASP-19b
obtained from the Spitzer satellite. 
We found good agreement between the observed
photometry and a spectrum based on a {\sc Drift-Marcs} model with $T_{\rm eff}$ = 2600\,K.

Gas giants in larger orbits will show more complicated structures due to their
more normal rotation speed, and will require more dynamic features and more 
extensive chemical gas and dust calculations included in the modeling due to 
their lower temperature. This will
be the subject of coming papers and more advanced versions of the {\sc Drift-Marcs}
code than presented here, as will the modeling of even more Earth-like exoplanets.
%Gas giants in larger orbits will show more complicated structures due to their
%more normal rotation speed, and will require more dynamic features and more 
%extensive chemical gas and dust calculations included in the modelling. This will
%be the subject of coming papers and more advanced versions of the DRIFT-MARCS
%code than presented here, as will the modelling of even more Earth-like exoplanets.

%-----------------------------------------------------------------------------------------------------------
% ACKNOWLEDGEMENTS
%-----------------------------------------------------------------------------------------------------------
\begin{acknowledgements}
We are thankful to University of St Andrews for hospitality and
financial support toward DJ during an extended stay in 2014 as part
of her PhD thesis work, where part of this work was done. ChH
highlight financial support of the European Community under the FP7
by the ERC starting grant 257431. 
We greatly appreciate discussions with, and valuable comments from, 
B. Gustafsson and K. Lodders{\color{black}, and are thankful for an 
inspiring and thorough referee report from D. Homeier}.
This research has benefitted from the SpeX Prism Spectral Libraries, maintained by Adam Burgasser 
at http://pono.ucsd.edu/~adam/browndwarfs/spexprism.
\end{acknowledgements}

%-----------------------------------------------------------------------------------------------------------
% BIBLIOGRAPHY
%-----------------------------------------------------------------------------------------------------------
\bibliographystyle{aa}
\bibliography{bibliography}

\begin{thebibliography}{115}
\expandafter\ifx\csname natexlab\endcsname\relax\def\natexlab#1{#1}\fi

\bibitem[{{Ackerman} \& {Marley}(2001)}]{ackerman01}
{Ackerman}, A.~S. \& {Marley}, M.~S. 2001, Astrophysical Journal, 556, 872

\bibitem[{{Allard} {et~al.}(2001){Allard}, {Hauschildt}, {Alexander},
  {Tamanai}, \& {Schweitzer}}]{allard01}
{Allard}, F., {Hauschildt}, P.~H., {Alexander}, D.~R., {Tamanai}, A., \&
  {Schweitzer}, A. 2001, Astrophysical Journal, 556, 357

\bibitem[{{Andersen} \& {J{\o}rgensen}(1995)}]{and1995}
{Andersen}, A.~C. \& {J{\o}rgensen}, U.~G. 1995, Meteoritics, 30

\bibitem[{{Anderson} {et~al.}(2010){Anderson}, {Gillon}, {Maxted}, {Barman},
  {Collier Cameron}, {Hellier}, {Queloz}, {Smalley}, \& {Triaud}}]{anderson10}
{Anderson}, D.~R., {Gillon}, M., {Maxted}, P.~F.~L., {et~al.} 2010, Astronomy
  \& Astrophysics, 513, L3

\bibitem[{{Anderson} {et~al.}(2013){Anderson}, {Smith}, {Madhusudhan},
  {Wheatley}, {Collier Cameron}, {Hellier}, {Campo}, {Gillon}, {Harrington},
  {Maxted}, {Pollacco}, {Queloz}, {Smalley}, {Triaud}, \& {West}}]{anderson13}
{Anderson}, D.~R., {Smith}, A.~M.~S., {Madhusudhan}, N., {et~al.} 2013, Monthly
  Notices of the RAS, 430, 3422

\bibitem[{{Aringer} {et~al.}(1997){Aringer}, {J{\o}rgensen}, \&
  {Langhoff}}]{ar1997}
{Aringer}, B., {J{\o}rgensen}, U.~G., \& {Langhoff}, S.~R. 1997, \aap, 323, 202

\bibitem[{{Aringer} {et~al.}(2002){Aringer}, {Kerschbaum}, \&
  {J{\o}rgensen}}]{ar2002}
{Aringer}, B., {Kerschbaum}, F., \& {J{\o}rgensen}, U.~G. 2002, \aap, 395, 915

\bibitem[{{Asplund} {et~al.}(2000){Asplund}, {Gustafsson}, {Lambert}, \&
  {Rao}}]{asp2000}
{Asplund}, M., {Gustafsson}, B., {Lambert}, D.~L., \& {Rao}, N.~K. 2000, \aap,
  353, 287

\bibitem[{{Barman} {et~al.}(2011){Barman}, {Macintosh}, {Konopacky}, \&
  {Marois}}]{barman11}
{Barman}, T.~S., {Macintosh}, B., {Konopacky}, Q.~M., \& {Marois}, C. 2011,
  Astrophysical Journal, 733, 65

\bibitem[{{Barton} {et~al.}(2013){Barton}, {Yurchenko}, \&
  {Jonathan}}]{barton13}
{Barton}, E.~J., {Yurchenko}, S.~N., \& {Jonathan}, T. 2013, Monthly Notices of
  the RAS, 434, 1469

\bibitem[{{Blackwell} {et~al.}(1995){Blackwell}, {Lynas-Gray}, \&
  {Smith}}]{black1995}
{Blackwell}, D.~E., {Lynas-Gray}, A.~E., \& {Smith}, G. 1995, \aap, 296, 217

\bibitem[{{Bohren} \& {Huffman}(1983)}]{bohren83}
{Bohren}, C.~F. \& {Huffman}, D.~R. 1983, {Absorption and Scattering of Light
  by Small Particles} ({John Wiley \& Sons Ltd})

\bibitem[{{Borysow} {et~al.}(2001){Borysow}, {J{\o}rgensen}, \&
  {Fu}}]{borysow2001}
{Borysow}, A., {J{\o}rgensen}, U.~G., \& {Fu}, Y. 2001, Journal of Quantitative
  Spectroscopy \& Radiative Transfer, 68, 235

\bibitem[{{Bowler} {et~al.}(2010){Bowler}, {Liu}, \& {Dupuy}}]{bowler10}
{Bowler}, B.~P., {Liu}, M.~C., \& {Dupuy}, T.~J. 2010, Astrophysical Journal,
  710, 45

\bibitem[{{Brooke} {et~al.}(2013){Brooke}, {Bernath}, {Schmidt}, \&
  {Bacskay}}]{brooke13}
{Brooke}, J. S.~A., {Bernath}, P.~F., {Schmidt}, T.~W., \& {Bacskay}, G.~B.
  2013, Journal of Quantitative Spectroscopy \& Radiative Transfer, 124, 11

\bibitem[{Brooke {et~al.}(2014{\natexlab{a}})Brooke, Bernath, Western, Hemert,
  \& Groenenboom}]{brooke14b}
Brooke, J. S.~A., Bernath, P.~F., Western, C.~M., Hemert, C.~v., \&
  Groenenboom, G.~C. 2014{\natexlab{a}}, The Journal of Chemical Physics, 141

\bibitem[{Brooke {et~al.}(2014{\natexlab{b}})Brooke, Ram, Western, Li,
  Schwenke, \& Bernath}]{brooke14}
Brooke, J. S.~A., Ram, R., Western, C.~M., {et~al.} 2014{\natexlab{b}},
  Astrophysical Journal Supplement Series, 210, 23

\bibitem[{{Bruggeman}(1935)}]{bruggeman35}
{Bruggeman}, D.~A.~G. 1935, Annalen der Physik, 416, 636

\bibitem[{{Burgasser}(2014)}]{burgasser14}
{Burgasser}, A.~J. 2014, in Astronomical Society of India Conference Series,
  Vol.~11, 7--16

\bibitem[{{Burgasser} {et~al.}(2009){Burgasser}, {Dhital}, \&
  {West}}]{burgasser09}
{Burgasser}, A.~J., {Dhital}, S., \& {West}, A.~A. 2009, Astronomical Journal,
  138, 1563

\bibitem[{{Burgasser} {et~al.}(2008){Burgasser}, {Liu}, {Ireland}, {Cruz}, \&
  {Dupuy}}]{burgasser08}
{Burgasser}, A.~J., {Liu}, M.~C., {Ireland}, M.~J., {Cruz}, K.~L., \& {Dupuy},
  T.~J. 2008, Astrophysical Journal, 681, 579

\bibitem[{{Burgasser} {et~al.}(2004){Burgasser}, {McElwain}, {Kirkpatrick},
  {Cruz}, {Tinney}, \& {Reid}}]{burgasser04}
{Burgasser}, A.~J., {McElwain}, M.~W., {Kirkpatrick}, J.~D., {et~al.} 2004,
  Astronomical Journal, 127, 2856

\bibitem[{Burrows {et~al.}(2005)Burrows, Dulick, Bauschlicher, Bernath, Ram,
  Sharp, \& Milsom}]{burrows05}
Burrows, A., Dulick, M., Bauschlicher, C.~W., {et~al.} 2005, Astrophysical
  Journal, 624, 988

\bibitem[{{Burrows} {et~al.}(1989){Burrows}, {Hubbard}, \&
  {Lunine}}]{burrows89}
{Burrows}, A., {Hubbard}, W.~B., \& {Lunine}, J.~I. 1989, Astrophysical
  Journal, 345, 939

\bibitem[{Burrows {et~al.}(2002)Burrows, Ram, Bernath, Sharp, \&
  Milsom}]{burrows02}
Burrows, A., Ram, R.~S., Bernath, P., Sharp, C.~M., \& Milsom, J.~A. 2002,
  Astrophysical Journal, 577, 986

\bibitem[{{Burrows} \& {Sharp}(1999)}]{burrows99}
{Burrows}, A. \& {Sharp}, C.~M. 1999, Astrophysical Journal, 512, 843

\bibitem[{{Burrows} {et~al.}(2006){Burrows}, {Sudarsky}, \&
  {Hubeny}}]{burrows06}
{Burrows}, A., {Sudarsky}, D., \& {Hubeny}, I. 2006, Astrophysical Journal,
  640, 1063

\bibitem[{{Carlson} {et~al.}(1988){Carlson}, {Rossow}, \& {Orton}}]{carlson88}
{Carlson}, B.~E., {Rossow}, W.~B., \& {Orton}, G.~S. 1988, Journal of
  Atmospheric Sciences, 45, 2066

\bibitem[{{Chiu} {et~al.}(2006){Chiu}, {Fan}, {Leggett}, {Golimowski}, {Zheng},
  {Geballe}, {Schneider}, \& {Brinkmann}}]{chiu06}
{Chiu}, K., {Fan}, X., {Leggett}, S.~K., {et~al.} 2006, Astronomical Journal,
  131, 2722

\bibitem[{{Cooper} {et~al.}(2003){Cooper}, {Sudarsky}, {Milsom}, {Lunine}, \&
  {Burrows}}]{cooper03}
{Cooper}, C.~S., {Sudarsky}, D., {Milsom}, J.~A., {Lunine}, J.~I., \&
  {Burrows}, A. 2003, Astrophysical Journal, 586, 1320

\bibitem[{Coppola {et~al.}(2011)Coppola, Lodi, \& Tennyson}]{coppola11}
Coppola, C.~M., Lodi, L., \& Tennyson, J. 2011, Monthly Notices of the RAS,
  415, 487

\bibitem[{{Decin} \& {Eriksson}(2007)}]{dec2007}
{Decin}, L. \& {Eriksson}, K. 2007, \aap, 472, 1041

\bibitem[{{Decin} {et~al.}(2003){Decin}, {Vandenbussche}, {Waelkens}, {Decin},
  {Eriksson}, {Gustafsson}, {Plez}, \& {Sauval}}]{dec2003}
{Decin}, L., {Vandenbussche}, B., {Waelkens}, C., {et~al.} 2003, \aap, 400, 709

\bibitem[{{Dominik} {et~al.}(1993){Dominik}, {Sedlmayr}, \& {Gail}}]{dominik93}
{Dominik}, C., {Sedlmayr}, E., \& {Gail}, H.-P. 1993, Astronomy \&
  Astrophysics, 277, 578

\bibitem[{Dorschner {et~al.}(1995)Dorschner, Begemann, Henning, J{\"a}ger, \&
  Mutschke}]{dorschner95}
Dorschner, J., Begemann, B., Henning, T., J{\"a}ger, C., \& Mutschke, H. 1995,
  Astronomy \& Astrophysics, 300, 503

\bibitem[{{Doughty} \& {Fraser}(1966)}]{doughty66b}
{Doughty}, N.~A. \& {Fraser}, P.~A. 1966, Monthly Notices of the RAS, 132, 267

\bibitem[{{Doughty} {et~al.}(1966){Doughty}, {Fraser}, \&
  {McEachran}}]{doughty66}
{Doughty}, N.~A., {Fraser}, P.~A., \& {McEachran}, R.~P. 1966, Monthly Notices
  of the RAS, 132, 255

\bibitem[{{Doyle}(1968)}]{doyle68}
{Doyle}, R.~O. 1968, Astrophysical Journal, 153, 987

\bibitem[{{Fortney} {et~al.}(2016){Fortney}, {Robinson}, {Domagal-Goldman},
  {Sk{\aa}lid Amundsen}, {Brogi}, {Claire}, {Crisp}, {Hebrard}, {Imanaka}, {de
  Kok}, {Marley}, {Teal}, {Barman}, {Bernath}, {Burrows}, {Charbonneau},
  {Freedman}, {Gelino}, {Helling}, {Heng}, {Jensen}, {Kane}, {Kempton},
  {Kopparapu}, {Lewis}, {Lopez-Morales}, {Lyons}, {Lyra}, {Meadows}, {Moses},
  {Pierrehumbert}, {Venot}, {Wang}, \& {Wright}}]{fort2016}
{Fortney}, J.~J., {Robinson}, T.~D., {Domagal-Goldman}, S., {et~al.} 2016,
  ArXiv e-prints

\bibitem[{{Gail} \& {Sedlmayr}(1988)}]{gail88}
{Gail}, H.-P. \& {Sedlmayr}, E. 1988, Astronomy \& Astrophysics, 206, 153

\bibitem[{GharibNezhad {et~al.}(2013)GharibNezhad, Shayesteh, \&
  Bernath}]{gharibnezhad13}
GharibNezhad, E., Shayesteh, A., \& Bernath, P. 2013, Monthly Notices of the
  RAS, 432, 2043

\bibitem[{{Gibson} {et~al.}(2010){Gibson}, {Aigrain}, {Pollacco}, {Barros},
  {Hebb}, {Hrudkov{\'a}}, {Simpson}, {Skillen}, \& {West}}]{gibson10}
{Gibson}, N.~P., {Aigrain}, S., {Pollacco}, D.~L., {et~al.} 2010, Monthly
  Notices of the RAS, 404, L114

\bibitem[{Grevesse {et~al.}(2007)Grevesse, Asplund, \& Sauval}]{grevesse07}
Grevesse, N., Asplund, M., \& Sauval, A.~J. 2007, Space Science Reviews, 130,
  105

\bibitem[{{Groh} {et~al.}(2013){Groh}, {Meynet}, {Georgy}, \&
  {Ekstr{\"o}m}}]{gro2013}
{Groh}, J.~H., {Meynet}, G., {Georgy}, C., \& {Ekstr{\"o}m}, S. 2013, \aap,
  558, A131

\bibitem[{{Gustafsson}(1971)}]{gu1971}
{Gustafsson}, B. 1971, \aap, 10, 187

\bibitem[{Gustafsson {et~al.}(1975)Gustafsson, Bell, Eriksson, \&
  Nordlund}]{gustafsson75}
Gustafsson, B., Bell, R.~A., Eriksson, K., \& Nordlund, A. 1975, Astronomy \&
  Astrophysics, 42, 407

\bibitem[{Gustafsson {et~al.}(2008)Gustafsson, Edvardsson, Eriksson,
  J{\o}rgensen, Nordlund, \& Plez}]{gustafsson08}
Gustafsson, B., Edvardsson, B., Eriksson, K., {et~al.} 2008, Astronomy \&
  Astrophysics, 486, 951

\bibitem[{Harris {et~al.}(2008)Harris, Larner, Tennyson, Kaminsky, Pavlenko, \&
  Jones}]{harris08}
Harris, G.~J., Larner, F.~C., Tennyson, J., {et~al.} 2008, Monthly Notices of
  the RAS, 390, 143

\bibitem[{Harris {et~al.}(2006)Harris, Tennyson, Kaminsky, Pavlenko, \&
  Jones}]{harris06}
Harris, G.~J., Tennyson, J., Kaminsky, B.~M., Pavlenko, Y.~C., \& Jones, H.
  R.~A. 2006, Monthly Notices of the RAS, 367, 400

\bibitem[{{Helling} {et~al.}(2008{\natexlab{a}}){Helling}, {Ackerman},
  {Allard}, {Dehn}, {Hauschildt}, {Homeier}, {Lodders}, {Marley}, {Rietmeijer},
  {Tsuji}, \& {Woitke}}]{helling08b}
{Helling}, C., {Ackerman}, A., {Allard}, F., {et~al.} 2008{\natexlab{a}},
  Monthly Notices of the RAS, 391, 1854

\bibitem[{{Helling} {et~al.}(1996){Helling}, {J{\o}rgensen}, {Plez}, \&
  {Johnson}}]{helling96}
{Helling}, C., {J{\o}rgensen}, U.~G., {Plez}, B., \& {Johnson}, H.~R. 1996,
  Astronomy \& Astrophysics, 315, 194

\bibitem[{{Helling} {et~al.}(2016){Helling}, {Lee}, {Dobbs-Dixon}, {Mayne},
  {Amundsen}, {Khaimova}, {Unger}, {Manners}, {Acreman}, \& {Smith}}]{hell2016}
{Helling}, C., {Lee}, G., {Dobbs-Dixon}, I., {et~al.} 2016, Monthly Notices of
  the RAS, 460, 855

\bibitem[{{Helling} {et~al.}(2001){Helling}, {Oevermann}, {L{\"u}ttke},
  {Klein}, \& {Sedlmayr}}]{helling01}
{Helling}, C., {Oevermann}, M., {L{\"u}ttke}, M.~J.~H., {Klein}, R., \&
  {Sedlmayr}, E. 2001, Astronomy \& Astrophysics, 376, 194

\bibitem[{Helling \& Woitke(2006)}]{helling06}
Helling, C. \& Woitke, P. 2006, Astronomy \& Astrophysics, 455, 325

\bibitem[{{Helling} {et~al.}(2008{\natexlab{b}}){Helling}, {Woitke}, \&
  {Thi}}]{helling08}
{Helling}, C., {Woitke}, P., \& {Thi}, W.-F. 2008{\natexlab{b}}, Astronomy \&
  Astrophysics, 485, 547

\bibitem[{{Hill} {et~al.}(2016){Hill}, {Christlieb}, {Beers}, {Barklem},
  {Kratz}, {Nordstr{\"o}m}, {Pfeiffer}, \& {Farouqi}}]{hill2016}
{Hill}, V., {Christlieb}, N., {Beers}, T.~C., {et~al.} 2016, ArXiv e-prints

\bibitem[{{H{\"o}fner} {et~al.}(1998){H{\"o}fner}, {J{\o}rgensen}, \&
  {Loidl}}]{ho1998}
{H{\"o}fner}, S., {J{\o}rgensen}, U.~G., \& {Loidl}, R. 1998, \apss, 255, 281

\bibitem[{{Hui-Bon-Hoa} {et~al.}(2000){Hui-Bon-Hoa}, {LeBlanc}, \&
  {Hauschildt}}]{hui-bon-hoa00}
{Hui-Bon-Hoa}, A., {LeBlanc}, F., \& {Hauschildt}, P.~H. 2000, Astrophysical
  Journal, 535, L43

\bibitem[{{Irwin}(1981)}]{irwin81}
{Irwin}, A.~W. 1981, Astrophysical Journal Supplement Series, 45, 621

\bibitem[{J{\"a}ger {et~al.}(2003)J{\"a}ger, Dorschner, Mutschke, Posch, \&
  Henning}]{jager03}
J{\"a}ger, C., Dorschner, J., Mutschke, H., Posch, T., \& Henning, T. 2003,
  Astronomy \& Astrophysics, 408, 193

\bibitem[{{John}(1967)}]{john67}
{John}, T.~L. 1967, Astrophysical Journal, 149, 449

\bibitem[{{J{\o}rgensen} {et~al.}(2000){J{\o}rgensen}, {Hammer}, {Borysow}, \&
  {Falkesgaard}}]{jor2000}
{J{\o}rgensen}, U.~G., {Hammer}, D., {Borysow}, A., \& {Falkesgaard}, J. 2000,
  \aap, 361, 283

\bibitem[{J{\o}rgensen {et~al.}(2001)J{\o}rgensen, Jensen, S{\o}rensen, \&
  Aringer}]{jorgensen01}
J{\o}rgensen, U.~G., Jensen, P., S{\o}rensen, G.~O., \& Aringer, B. 2001,
  Astronomy \& Astrophysics, 372, 249

\bibitem[{{J{\o}rgensen} {et~al.}(1992){J{\o}rgensen}, {Johnson}, \&
  {Nordlund}}]{jorgensen92}
{J{\o}rgensen}, U.~G., {Johnson}, H.~R., \& {Nordlund}, A. 1992, \aap, 261, 263

\bibitem[{{Karzas} \& {Latter}(1961)}]{karzas61}
{Karzas}, W.~J. \& {Latter}, R. 1961, Astrophysical Journal Supplement Series,
  6, 167

\bibitem[{{Kirkpatrick} {et~al.}(2010){Kirkpatrick}, {Looper}, {Burgasser},
  {Schurr}, {Cutri}, {Cushing}, {Cruz}, {Sweet}, {Knapp}, {Barman},
  {Bochanski}, {Roellig}, {McLean}, {McGovern}, \& {Rice}}]{kirkpatrick10}
{Kirkpatrick}, J.~D., {Looper}, D.~L., {Burgasser}, A.~J., {et~al.} 2010,
  Astrophysical Journal Supplement, 190, 100

\bibitem[{Kupka {et~al.}(2011)Kupka, Dubernet, \& Collaboration}]{kupka11}
Kupka, F., Dubernet, M.-L., \& Collaboration, V. 2011, Baltic Astronomy, 20,
  503

\bibitem[{{Kurucz}(1970)}]{kurucz70}
{Kurucz}, R.~L. 1970, SAO Special Report, 309

\bibitem[{{Kurucz}(2011)}]{kurucz11}
{Kurucz}, R.~L. 2011, Canadian Journal of Physics, 89, 417

\bibitem[{{Lambert} {et~al.}(1986){Lambert}, {Gustafsson}, {Eriksson}, \&
  {Hinkle}}]{lambert86}
{Lambert}, D.~L., {Gustafsson}, B., {Eriksson}, K., \& {Hinkle}, K.~H. 1986,
  Astrophysical Journal Supplement Series, 62, 373

\bibitem[{{Lee} {et~al.}(2015){Lee}, {Helling}, {Dobbs-Dixon}, \&
  {Juncher}}]{lee2015}
{Lee}, G., {Helling}, C., {Dobbs-Dixon}, I., \& {Juncher}, D. 2015, \aap, 580,
  A12

\bibitem[{{Lewis}(1969)}]{lewis69}
{Lewis}, J.~S. 1969, Icarus, 10, 365

\bibitem[{{Ludwig} {et~al.}(2002){Ludwig}, {Allard}, \&
  {Hauschildt}}]{ludwig02}
{Ludwig}, H.-G., {Allard}, F., \& {Hauschildt}, P.~H. 2002, Astronomy \&
  Astrophysics, 395, 99

\bibitem[{{Lunine} {et~al.}(1986){Lunine}, {Hubbard}, \& {Marley}}]{lunine86}
{Lunine}, J.~I., {Hubbard}, W.~B., \& {Marley}, M.~S. 1986, Astrophysical
  Journal, 310, 238

\bibitem[{{Madhusudhan} \& {Seager}(2009)}]{madhusudhan09}
{Madhusudhan}, N. \& {Seager}, S. 2009, Astrophysical Journal, 707, 24

\bibitem[{{Madhusudhan} \& {Seager}(2010)}]{madhusudhan10}
{Madhusudhan}, N. \& {Seager}, S. 2010, Astrophysical Journal, 725, 261

\bibitem[{{Madhusudhan} \& {Seager}(2011)}]{madhusudhan11}
{Madhusudhan}, N. \& {Seager}, S. 2011, Astrophysical Journal, 729, 41

\bibitem[{{Markwardt}(2009)}]{markwardt09}
{Markwardt}, C.~B. 2009, in Astronomical Society of the Pacific Conference
  Series, Vol. 411, Astronomical Data Analysis Software and Systems XVIII, 251

\bibitem[{Masseron {et~al.}(2014)Masseron, Plez, Eck, Colin, Daoutidis,
  Godefroid, Coheur, Bernath, Jorissen, \& Christlieb}]{masseron14}
Masseron, T., Plez, B., Eck, S.~V., {et~al.} 2014, Astronomy \& Astrophysics,
  571, A47

\bibitem[{{Matrozis} {et~al.}(2013){Matrozis}, {Ryde}, \& {Dupree}}]{mat2013}
{Matrozis}, E., {Ryde}, N., \& {Dupree}, A.~K. 2013, \aap, 559, A115

\bibitem[{Mie(1908)}]{mie08}
Mie, G. 1908, Annalen der Physik, 330, 377

\bibitem[{{Mihalas}(1965)}]{mihalas65}
{Mihalas}, D. 1965, Astrophysical Journal Supplement,, 9, 321

\bibitem[{{Mihalas}(1978)}]{mihalas78}
{Mihalas}, D. 1978, {Stellar atmospheres} (W.~H.~Freeman and Co.)

\bibitem[{{Nissen} {et~al.}(2014){Nissen}, {Chen}, {Carigi}, {Schuster}, \&
  {Zhao}}]{nis2014}
{Nissen}, P.~E., {Chen}, Y.~Q., {Carigi}, L., {Schuster}, W.~J., \& {Zhao}, G.
  2014, \aap, 568, A25

\bibitem[{{Palik}(1985)}]{palik85}
{Palik}, E.~D. 1985, {Handbook of Optical Constants of Solids} ({Academic
  Press})

\bibitem[{{Peach}(1970)}]{peach70}
{Peach}, G. 1970, Memoirs of the RAS, 73, 1

\bibitem[{{Plez}(1992)}]{plez92}
{Plez}, B. 1992, Astrophysical Journal Supplement Series, 94, 527

\bibitem[{{Plez} {et~al.}(2003){Plez}, {van Eck}, {Jorissen}, {Edvardsson},
  {Eriksson}, \& {Gustafsson}}]{plez03}
{Plez}, B., {van Eck}, S., {Jorissen}, A., {et~al.} 2003, in IAU Symposium,
  Vol. 210, Modelling of Stellar Atmospheres, ed. N.~{Piskunov}, W.~W. {Weiss},
  \& D.~F. {Gray}, 2P

\bibitem[{Popovas(2014)}]{popovas14}
Popovas, A. 2014, Master's thesis, University of Copenhagen

\bibitem[{Posch {et~al.}(2003)Posch, Kerschbaum, Fabian, Mutschke, Dorschner,
  Tamanai, \& Henning}]{posch03}
Posch, T., Kerschbaum, F., Fabian, D., {et~al.} 2003, Astrophysical Journal
  Supplement Series, 149, 437

\bibitem[{{Rayner} {et~al.}(2003){Rayner}, {Toomey}, {Onaka}, {Denault},
  {Stahlberger}, {Vacca}, {Cushing}, \& {Wang}}]{rayner03}
{Rayner}, J.~T., {Toomey}, D.~W., {Onaka}, P.~M., {et~al.} 2003, The
  Publications of the Astronomical Society of the Pacific, 115, 362

\bibitem[{{Rossow}(1978)}]{rossow78}
{Rossow}, W.~B. 1978, Icarus, 36, 1

\bibitem[{Rothman {et~al.}(2010)Rothman, Gordon, Barber, Dothe, Gamache,
  Goldman, Perevalov, Tashkun, \& Tennyson}]{rothman10}
Rothman, L.~S., Gordon, I.~E., Barber, R.~J., {et~al.} 2010, Journal of
  Quantitative Spectroscopy \& Radiative Trasnfer, 111, 2139

\bibitem[{{Ryde} {et~al.}(2002){Ryde}, {Lambert}, {Richter}, \&
  {Lacy}}]{ry2002}
{Ryde}, N., {Lambert}, D.~L., {Richter}, M.~J., \& {Lacy}, J.~H. 2002, \apj,
  580, 447

\bibitem[{{Schmidt} {et~al.}(2010){Schmidt}, {West}, {Burgasser}, {Bochanski},
  \& {Hawley}}]{schmidt10}
{Schmidt}, S.~J., {West}, A.~A., {Burgasser}, A.~J., {Bochanski}, J.~J., \&
  {Hawley}, S.~L. 2010, Astronomical Journal, 139, 1045

\bibitem[{Schwenke(1998)}]{schwenke98}
Schwenke, D.~W. 1998, Faraday Discussions, 109, 321

\bibitem[{{Sharp} \& {Huebner}(1990)}]{sharp90}
{Sharp}, C.~M. \& {Huebner}, W.~F. 1990, Astrophysical Journal Supplement
  Series, 72, 417

\bibitem[{{Siqueira-Mello} {et~al.}(2016){Siqueira-Mello}, {Chiappini},
  {Barbuy}, {Freeman}, {Ness}, {Depagne}, {Cantelli}, {Pignatari}, {Hirschi},
  {Frischknecht}, {Meynet}, \& {Maeder}}]{siq2016}
{Siqueira-Mello}, C., {Chiappini}, C., {Barbuy}, B., {et~al.} 2016, ArXiv
  e-prints

\bibitem[{{Somerville}(1964)}]{somerville64}
{Somerville}, W.~B. 1964, Astrophysical Journal, 139, 192

\bibitem[{{Somerville}(1965)}]{somerville65}
{Somerville}, W.~B. 1965, Astrophysical Journal, 141, 811

\bibitem[{{Street} {et~al.}(2015){Street}, {Fulton}, {Scholz}, {Horne},
  {Helling}, {Juncher}, {Lee}, \& {Valenti}}]{street15}
{Street}, R.~A., {Fulton}, B.~J., {Scholz}, A., {et~al.} 2015, Astrophysical
  Journal, 812, 161

\bibitem[{{Tsuji}(1964)}]{tsuji64}
{Tsuji}, T. 1964, Annals of the Tokyo Astronomical Observatory, 9, 1

\bibitem[{Tsuji(1973)}]{tsuji73}
Tsuji, T. 1973, Astronomy \& Astrophysics, 23, 411

\bibitem[{{Tsuji}(2001)}]{tsuji01}
{Tsuji}, T. 2001, in Ultracool Dwarfs: New Spectral Types L and T, ed. H.~R.~A.
  {Jones} \& I.~A. {Steele}, 9

\bibitem[{{Tsuji} {et~al.}(1996){Tsuji}, {Ohnaka}, {Aoki}, \&
  {Nakajima}}]{tsuji96b}
{Tsuji}, T., {Ohnaka}, K., {Aoki}, W., \& {Nakajima}, T. 1996, Astronomy \&
  Astrophysics, 308, L29

\bibitem[{{Van Eck} {et~al.}(2017){Van Eck}, {Neyskens}, {Jorissen}, {Plez},
  {Edvardsson}, {Eriksson}, {Gustafsson}, {J{\o}rgensen}, \&
  {Nordlund}}]{vaneck16}
{Van Eck}, S., {Neyskens}, P., {Jorissen}, A., {et~al.} 2017, \aap, 601, A10

\bibitem[{{Walker}(2014)}]{walker14}
{Walker}, R. 2014,
  \url{http://www.ursusmajor.ch/downloads/spectroscopic-atlas-5_0-english.pdf}

\bibitem[{Weck {et~al.}(2003)Weck, Stancil, \& Kirby}]{weck03}
Weck, P.~F., Stancil, P.~C., \& Kirby, K. 2003, The Journal of Chemical
  Physics, 118, 9997

\bibitem[{Wende {et~al.}(2010)Wende, Reiners, Seifahrt, \& Bernath}]{wende10}
Wende, S., Reiners, A., Seifahrt, A., \& Bernath, P.~F. 2010, Astronomy \&
  Astrophysics, 523, A58

\bibitem[{{Witte} {et~al.}(2011){Witte}, {Helling}, {Barman}, {Heidrich}, \&
  {Hauschildt}}]{witte2011}
{Witte}, S., {Helling}, C., {Barman}, T., {Heidrich}, N., \& {Hauschildt},
  P.~H. 2011, \aap, 529, A44

\bibitem[{{Witte} {et~al.}(2009){Witte}, {Helling}, \& {Hauschildt}}]{witte09}
{Witte}, S., {Helling}, C., \& {Hauschildt}, P.~H. 2009, Astronomy \&
  Astrophysics, 506, 1367

\bibitem[{Woitke \& Helling(2003)}]{woitke03}
Woitke, P. \& Helling, C. 2003, Astronomy \& Astrophysics, 399, 297

\bibitem[{Woitke \& Helling(2004)}]{woitke04}
Woitke, P. \& Helling, C. 2004, Astronomy \& Astrophysics, 414, 335

\bibitem[{Yadin {et~al.}(2012)Yadin, Veness, Conti, Hill, Yurchnko, \&
  Tennyson}]{yadin12}
Yadin, B., Veness, T., Conti, P., {et~al.} 2012, Monthly Notices of the RAS,
  425, 34

\bibitem[{Zeidler {et~al.}(2013)Zeidler, Posch, \& Mutschke}]{zeidler13}
Zeidler, S., Posch, T., \& Mutschke, H. 2013, Astronomy \& Astrophysics, 553,
  A81

\end{thebibliography}

%-----------------------------------------------------------------------------------------------------------
% APPENDIX
%-----------------------------------------------------------------------------------------------------------
\appendix

\section{Included atoms and molecules}\label{s:atoms_and_molecules}
The following atoms and molecules were included in the chemical equilibrium calculations in {\sc Marcs}. \\\\
{\bf Atoms (38):}\\
H, He, Li, Be, B, C, N, O, F, Na, Mg, Al, Si, P, S, Cl, K, Ca, Sc, Ti, V, Cr, Mn, Fe, Ni, Cu, Ge, 
Br, Rb, Sr, Y, Zr, Nb, I, Ba, La, Ce, Nd.\\\\
{\bf Molecules (210):}\\
H$^-$, H$_2$, H$_2$O, OH, CH, CO, CN, C$_2$, N$_2$, O$_2$, NO, NH, C$_2$H$_2$, 
HCN, C$_2$H, HS, SiH, C$_3$, CS, SiC, SiC$_2$, NS, SiN, SiO, SO, S$_2$, SiS, TiO, VO, 
ZrO, MgH, HF, HCl, CH$_4$, CH$_2$, CH$_3$, NH$_2$, NH$_3$, C$_2$N$_2$, C$_2$N, 
CO$_2$, F$^-$, AlF, CaF, CaF$_2$, MgOH, Al$_2$O, AlOH, AlOF, AlOCl, NaOH, Si$_2$C, 
SiO$_2$, H$_2$S, CS$_2$, AlCl, NaCl, KCl, KOH, CaCl, CaCl$_2$, CaOH, TiO$_2$, VO$_2$, 
LiH, LiO, LiF, LiCl, BeH$_2$, BeO, BeF, BeCl, BeCl$_2$, BeOH, BH, BH$_2$, BO, B$_2$O, 
BS, BF, BCl, HBO, HBO$_2$, C$^-$, C$_2^-$, C$_2$H$_4$, NO$_2^-$, N$_2$H$_2$, 
N$_2$H$_4$, CN$_2$, C$_4$N$_2$, NO$_2$, NO$_3$, N$_2$O, N$_2$O$_4$, HNO, 
HNO$_2$, HNO$_3$, HCNO, O$^-$, O$_2^-$, OH$^-$, CO$_2^-$, C$_2$O, HCO, H$_2$CO, 
F$_2$, FO, NaH, NaO, NaF, MgO, MgS, MgF, MgF$_2$, MgCl, MgCl$_2$, AlH, AlO, AlO$_2$, 
AlS, AlF$_2$, AlCl$_2$, SI$^-$, SiH$_4$, SiF, SiF$_2$, SiCl, SiCl$_2$, PH, PH$_2$, PH$_3$, 
CP, NP, PO, PO$_2$, PS, PF, PF$_2$, PCL, COS, SO$_2$, S$_2$O, SO$_3$, Cl$^-$, Cl$_2$, 
CCl, CCl$_2$, CCl$_3$, CCl$_4$, ClO, ClO$_2$, Cl$_2$O, SCl, SCl$_2$, HClO, CClO, KH, 
KO, KF, CaO, CaS, TiF, TiF$_2$, TiCl, TiCl$_2$, VN, CrN, CrO, CrO$_2$, FeO, FeS, FeF, 
FeF$_2$, FeCl, FeCl$_2$, NiCl, CuO, CuF, CuCl, SrO, SrS, SrF, SrF$_2$, SrCl, SrCl$_2$, 
SrOH, ZrH, ZrN, ZrO$_2$, ZrF, ZrF$_2$, ZrCl, ZrCl$_2$, HI, BaO, BaS, BaF, BaF$_2$, BaCl, 
BaCl$_2$, BaOH, NBO, C$_4$, C$_5$, TiH, CaH, FeH, CrH.

\section{Synthetic spectra decomposition}\label{s:detailed_spectrum}
\begin{figure*}
\includegraphics[width=0.95\hsize]{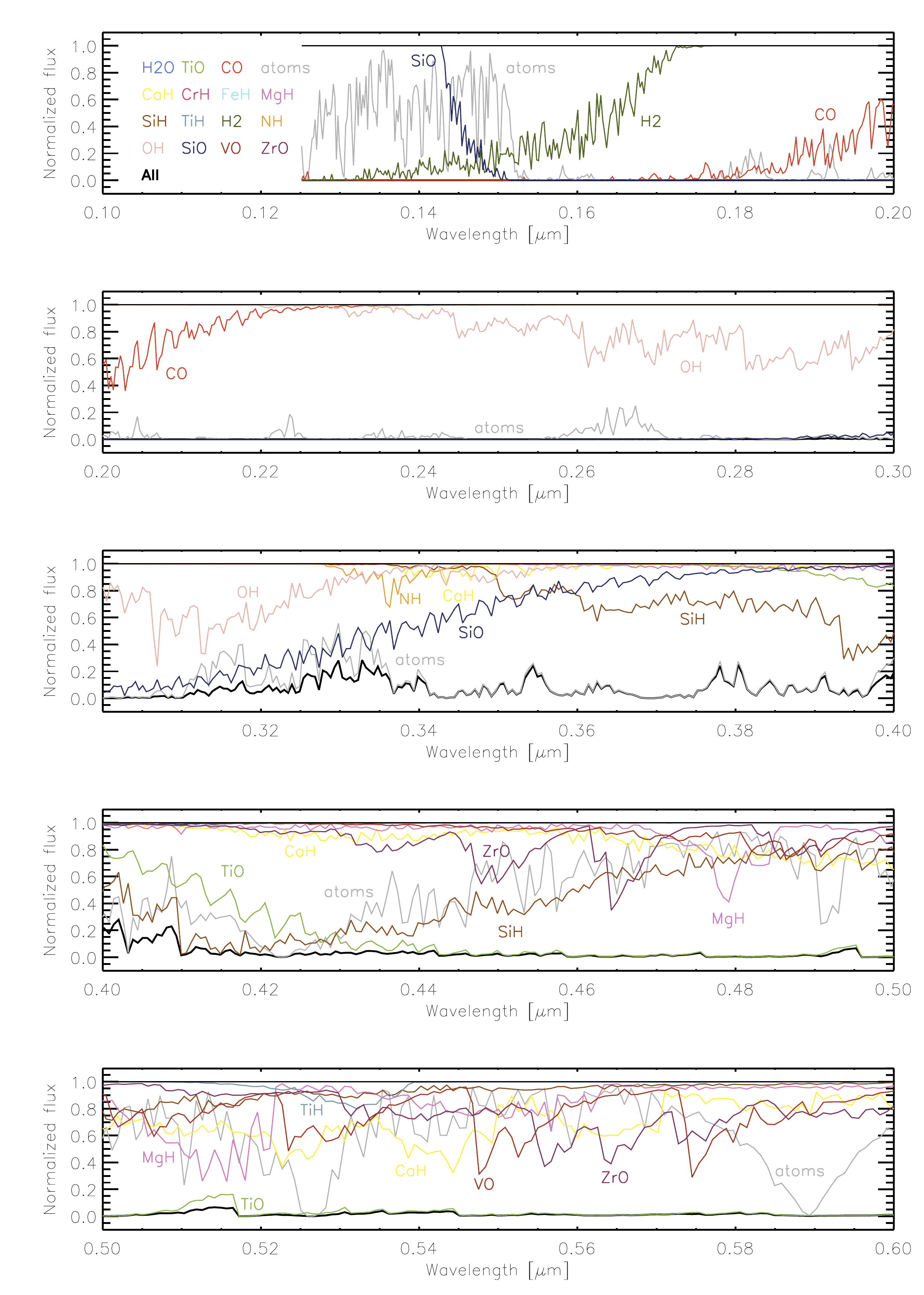}
{\ }\\*[-0.7cm]
\caption{Spectral contributions of atomic and molecular opacity sources for a cloud-free {\sc Marcs}-model atmosphere (of $T_{\rm eff}$=3000 K, $\log(g)$=4.5, solar element abundances) for the spectral range $\lambda=0.1-0.6\ \mu$m (top to bottom panel).} 
\label{fig:specfree1}
\end{figure*}

\begin{figure*}
\includegraphics[width=0.95\hsize]{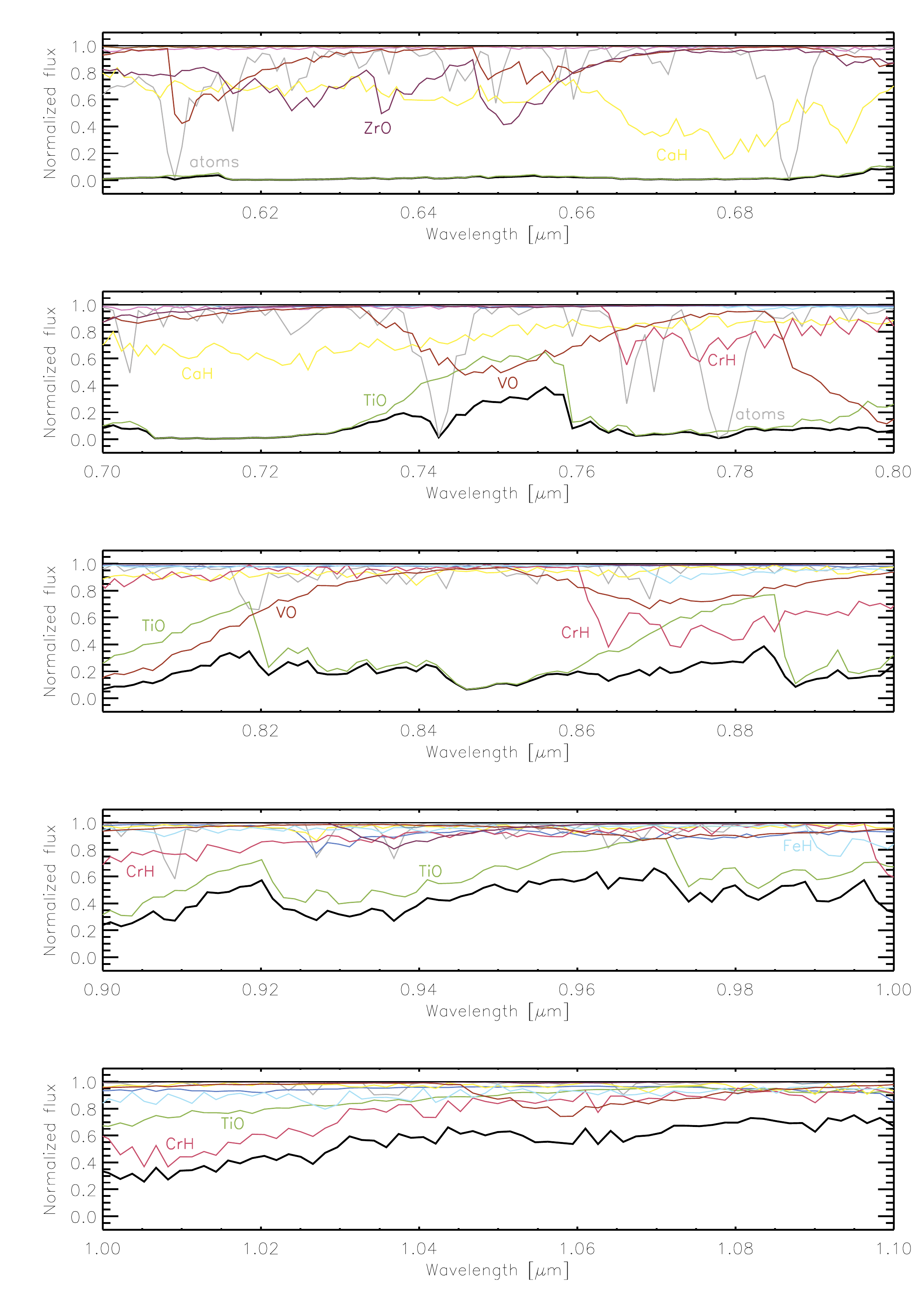}
{\ }\\*[-0.7cm]
\caption{Same as Figure~\ref{fig:specfree1} but for $\lambda=0.6-1.1\ \mu$m (top to bottom panel).} 
\label{fig:specfree2}
\end{figure*}

\begin{figure*}
\includegraphics[width=0.95\hsize]{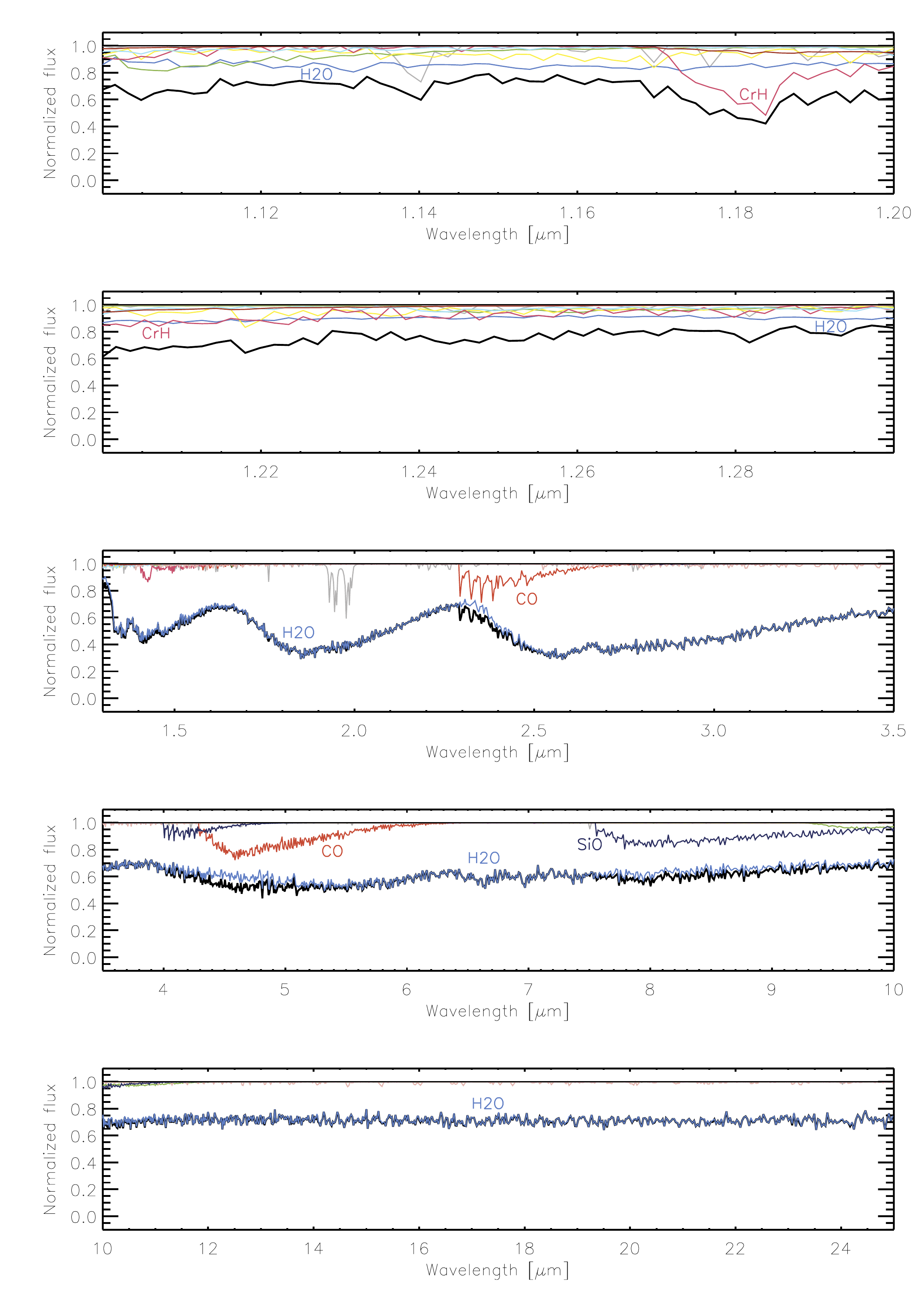}
{\ }\\*[-0.7cm]
\caption{Same as Figure~\ref{fig:specfree1} but for $\lambda=1.1-24\ \mu$m (top to bottom panel).} 
\label{fig:specfree3}
\end{figure*}

We provide a detailed decomposition of the gas-contributions in Figs.~\ref{fig:specfree1}--\ref{fig:specfree3}.
At the shortest wavelengths, SiO, H$_2$, and CO are all very strong absorbers, with
 SiO being the most influential from $1.8 - 3$ $\mu$m. OH also
 has a fairly strong absorption from $2.6 - 3.2$ $\mu$m, but it
 is obscured by the SiO absorption. NH makes a short appearance around
 3.4 $\mu$m. TiO absorption starts to grow from 4 $\mu$m and completely
 dominates the spectrum from $4.4 - 9$ $\mu$m, with a few
 exceptions; at 7.5 $\mu$m, and 8.8 $\mu$m, the absorption of TiO
 weakens but is compensated for by the absorption of VO and CrH,
 respectively.  In fact, if there had been no TiO in the atmosphere,
 the metallic hydrides would have provided most of the absorption from
 $0.4 - 1.1\ \mu$m with CaH peaking at 0.68 $\mu$m, CrH at 0.88
 $\mu$m and 1 $\mu$m, FeH at 1 $\mu$m, MgH at 0.51 $\mu$m, SiH at
 0.42 $\mu$m and TiH at 0.53 $\mu$m.  ZrO also shows its strongest
 absorption in this region. Finally, H$_2$O absorption shows up at
 1.1 $\mu$m and completely dominates the spectrum in the infrared and
 beyond.  LiH and NO absorption both have a negligible effect on the
 spectrum because of their very low partial pressures. Even though the
 absorption coefficient of CO$_2$ is larger than that of CO in the
 optical, the partial pressure of CO$_2$ at these high temperatures is
 less than a thousandth of the partial pressure of CO, and its
 spectroscopic features are therefore almost imperceptible.  CH,
 C$_2$, CN and HCN are barely present in oxygen-rich atmospheres at
 these temperatures, and their contribution to the absorption is
 consequently negligible. These molecules will, however, become of
 interest if carbon is enhanced compared to the solar C/O ratio. 


\end{document}